\documentclass{elsarticle}

\usepackage{etoolbox}
\makeatletter
\patchcmd{\ps@pprintTitle}{\footnotesize\itshape
       Preprint submitted to \ifx\@journal\@empty Elsevier
       \else\@journal\fi\hfill\today}{\relax}{}{}
\makeatother

\usepackage{fancyhdr}

\fancypagestyle{specialfooter}{%
  \fancyhf{}
  
  \fancyfoot[L]{{ \fontsize{7}{7} \selectfont  
  Received 8 January, Revised 21 May 2019, Accepted 9 June 2019, Available online 3 July 2019. 
  \\ \medskip
  \textbf{Published in Information Processing \& Management, Elsevier, Volume 56, Issue 6. }
  \\
  \textbf{Published version available at https://doi.org/10.1016/j.ipm.2019.102055.sy} 
  \\ \medskip
  @ 2019. This manuscript version is made available under the CC-BY-NC-ND 4.0 license
  (http://creativecommons.org/licenses/by-nc-nd/4.0/)
 }
}}

\usepackage{lingmacros}
\usepackage{tree-dvips}

\usepackage[utf8]{inputenc}
\usepackage{svg}
\usepackage{multirow}

\usepackage{mathtools}
\journal{Information Processing \& Management}

\begin{document}

\begin{frontmatter}

\title{The Evolution of Argumentation Mining: From Models to Social Media and Emerging Tools}

\author[add1]{Anastasios Lytos}
\ead{alytos1@sheffield.ac.uk}
\author[add2,add3]{Thomas Lagkas}
\ead{t.lagkas@sheffield.ac.uk}
\author[add4]{Panagiotis Sarigiannidis}
\ead{psarigiannidis@uowm.gr}
\author[add1]{Kalina Bontcheva}
\ead{k.bontcheva@sheffield.ac.uk}

\address[add1]{Department of Computer Science, The University of Sheffield, Sheffield, UK}
\address[add2]{Computer Science Department, The University of Sheffield International Faculty - CITY
College, Thessaloniki, Greece}
\address[add3]{South East European Research Centre (SEERC), Thessaloniki, Greece}
\address[add4]{Department of Informatics and Telecommunications Engineering, University of Western
Macedonia, Kozani, Greece}


\begin{abstract}
Argumentation mining is a rising subject in the computational linguistics domain focusing on extracting structured arguments from natural text, often from unstructured or noisy text.  The initial approaches on modeling arguments was aiming to identify a flawless argument on specific fields (Law, Scientific Papers) serving specific needs (completeness, effectiveness).  With the emerge of Web 2.0 and the explosion in the use of social media both the diffusion of the data and the argument structure have changed.  In this survey article, we bridge the gap between theoretical approaches of argumentation mining and pragmatic schemes that satisfy the needs of social media generated data, recognizing the need for adapting more flexible and expandable schemes, capable to adjust to the argumentation conditions that exist in social media. We review, compare, and classify existing approaches, techniques and tools, identifying the positive outcome of combining tasks and features, and eventually propose a conceptual architecture framework.  The proposed theoretical framework is an argumentation mining scheme able to identify the distinct sub-tasks and capture the needs of social media text, revealing the need for adopting more flexible and extensible frameworks.

\end{abstract}


\end{frontmatter}


\section{Introduction}
\label{S:Introduction}

\thispagestyle{specialfooter}

Argumentation mining (AM) is identified as a multidisciplinary research topic, with roots on rhetoric and philosophy \cite{Toulmin2003TheArgument}, which gained the interest of the scientific community because of its potential when novel Artificial Intelligence (AI) algorithms and techniques are exploited. The recent advances in Machine Learning (ML) in combination with the emergence of social web can enable impressive progress in different scientific fields with great impact on commercial applications. An AM system has the capacity to mine and analyse a great volume of text data through a variety of sources, providing tools for policy-making and socio-political sciences \cite{Liebeck2016WhatFeld,Addawood2016WhatMedia,Boltuzic2014BackDiscussions}, software engineering \cite{Kurtanovic2018OnEngineering}, while it opens new horizons for the broader area of business, economics and finance, with the digital marketing being the most promising field \cite{Park2014IdentifyingComments,Park2015ConditionalComments,Rajendran2016ContextualReviews,Rajendran2018IsDataset}. 

Opinion mining and sentiment analysis could be characterized as the predecessors of AM in a simplified form and their limits have already been questioned in the effort for seeking a deeper understanding of the human reasoning \cite{Belbachir2018UsingDetection, Tubishat2018ImplicitChallenges}. As AM we define a series of actions that could be independent or connected to each other and they are relevant to the tasks of detection, extraction and evaluation of arguments, where argument is a piece of text offering evidence or reasoning in favor or against a specific topic. 

The core of the human reasoning is the argumentation process and AM-related tasks attempt to solve a series of problems such as the detection of an argumentative stance towards a specific object, the analysis and evaluation of the argument's components and the detection of possible relations between them. The cohesion in the components of the arguments and the existence of backing in the claims is a major objective in argument mining because these characteristics have the ability to alter a claim to an argument. For human beings the interpretation of an argument is realized instantly without needing any special effort because of their skill to grasp the context of the information and to accomplish connections with previous experiences (facts, opinions, feelings).  

This ability of combining multiple sources of information is missing in existing argument models as it is difficult even for human annotators to agree upon specific guidelines for the modeling of an argument. Furthermore, identifying the thin line between implicit and explicit arguments is harsh as often annotators are implicitly driven by the context. Probably that is also the reason why the majority of research is focused on structured data like law text \cite{Mochales2011ArgumentationMining,Savelka2016ExtractingTerms,Lippi2018CLAUDETTE:Service}, scientific text \cite{Green2018TowardsSchemes, Lauscher2018ArguminSci:Writing, Lauscher2018AnPublications}, formal debates \cite{Naderi2015ArgumentationDiscourse} or news articles \cite{Bal2010TowardsEditorials,Sardianos2015ArgumentNews} instead of unstructured text like informal discourse and web-user generated data, although recently there have been some notable research endeavors towards this direction, which are presented the section \ref{sec:AMSocialMedia}.

Annotating and automatically analysing arguments from Web with great heterogeneity of contents and diversity of jargon is a challenging task. Arguments in social media and informal discourse are sometimes implicit, meaning that the logical structure of an argument's components (premises, claims, warrants) are not always spelled out and instantly distinguishable, hence, analysis must take place in order to determine the distinctive components. In text derived from social media arguments frequently are missing, as it is common a tweet or a facebook post to contain just a stance on a specific topic without supporting it with evidence or reasoning. 

The decoding of the human reasoning process into computer language is a challenging task as it consists of many subprocesses that are difficult to be separated and analysed. The medium for arguing for human beings is natural language, whereas the input for ML algorithms and techniques should be distinct, structured and composed with well-established rules. A wide range of methodologies have been applied for modeling natural language, such as explicit distinctive components \cite{Toulmin2003TheArgument}, argumentative zoning \cite{REED2007ArgumentIntelligence}, tree structures, dialog-oriented diagrams \cite{Skeppstedt2016UnsharedDebates}, serial structure of arguments \cite{Freeman2011ArgumentTheory,Walton2011HowIntelligence} and modifications to simpler structures of existing schemes \cite{Peldszus2016RhetoricalText,Stab2014AnnotatingEssays}.

However, the claim or other parts of an argument might be implicit \cite{Green2017ManualMining,Rajendran2016ContextualReviews,Boltuzic2016FillDebates,Habernal2018TheWarrants,Rajendran2018IsDataset} and tacit assumptions or premises (enthymemes) take place related to commonsense reasoning. This process is named completion or enthymematic argumentation and takes place often and unconsciously in casual discourse that can be found in Web-generated data. The distinction between explicit and implicit argument lies on the presence of certain syntactic constructions or lexemes (such as conjunctions). Implicit arguments, where lack of these characteristics is noticed, can be identified through previously gained knowledge and logical inference.

This a-priori knowledge is extremely difficult to be expressed through conventional argumentation schemes, which demand a strict structure of the components of the argument. The early approaches \cite{Toulmin2003TheArgument,MANN1988RhetoricalOrganization} were focusing on the philosophical aspect of the argument, whereas modern approaches consider unstructured data and implicit relations between the components of the argument \cite{Green2017ManualMining,Rajendran2016ContextualReviews,Boltuzic2016FillDebates,Habernal2018TheWarrants,Rajendran2018IsDataset}. For example, in Toulmin's model \cite{Toulmin2003TheArgument,Reisert2015AArgumentation} (figure \ref{fig:ToulminExample}), which has a great impact on modern argumentation schemes, a detailed microstructure is proposed with six specified components, 1) an indisputable \textit{datum}, 2) a subjective \textit{claim} on the foundation of datum, 3) a \textit{warrant} that links them imposed by logical inference, 4) the \textit{backing} of the justification 5) leading to a degree of confidence \textit{(qualifiers)} as long as 6) a \textit{rebuttal} cannot withstand the claim. Data that do not have a structured, well-specified format are hard to be represented by such a strict model.

Breaking down arguments deriving from web or from casual discourse is a demanding task with doubts if it is even feasible. An evidence of the previous statement is the fact that the field of opinion mining thrives in social media data \cite{Lee2018UnderstandingFacebook, Nguyen2018MultilingualEmbedding} and especially in twitter \cite{ChandraPandey2017TwitterMethod, Giachanou2016LikeNot}, on the contrary only limited research has been conducted on AM in unstructured data and fewer frameworks have been designed able to capture the special features of social media. In 5.1 we propose a conceptual framework able to capture the specific features of social media text and also enhance other tasks in the wider area of NLP with the use of argumentative features.  

There is a small number of related works surveying argumentation mining. Peldszus and Stede \cite{Peldszus2013FromTexts} realized a thorough revision of models and diagrams for arguments, able to be exploited in the context of AM. We provide the necessary background for argumentation schemes and we evaluate them based on their suitability on social media text, but we do not devote the entire paper to this topic. Both \cite{Lippi2016ArgumentationMining} and \cite{Cabrio2018FiveAnalysis}, surveyed a big spectrum of the AM field, describing models, corpora and methods, but they overlook the special nature of social media and the special features they present. On contrary, we focus on text derived from social media and our entire approach is based on the characteristics of the social media; chaotic nature, noisy text, vague claims, complicated network relations, implicit premises, etc. While the authors in \cite{Habernal2017ArgumentationDiscourse} have addressed AM in web-generated data, they provided limited information about the connection of AM with its distinctive sub-tasks or other NLP tasks. Instead, we perceive AM as an entire pipeline with distinctive sub-tasks, able to be both stand-alone and correlate with each other, and therefore boost existing tasks such as sentiment analysis, topic modeling, rumour identification, etc.

Our contribution lies on the following main axes: bridge the gap between areas that are interrelated, but without having explicit connection, illustrate the current methods, tasks and tools that exist in the wider area of AM, and stress the importance of automatic AM mechanisms and tools. This extensive survey paper illustrates the evolution of argumentation and reveals the formation of AM as a scientific field through a complete review of its roots and the needs that currently form it, followed by a thorough presentation and comparison of the existing AM approaches in text derived from social media. The existing tools for the wider area of NLP are presented, assessing their potential use in AM, followed by the presentation of the tools that have been specifically built for argumentation. Furthermore, a complete conceptual architecture for AM is proposed, covering its different subtasks and possible relations with other fields.

In the next section, the most influential logical schemes are presented as well as the first attempts of connecting the argumentation process with AI. Every aspect of the AM problem is presented in section \ref{sec:AMSocialMedia} with a focus on noisy data derived from social media. Section \ref{sec:Tools} presents the existing tools in the field of AM and classifies them into two categories: general-purpose NLP tools and tools specifically designed for the task of AM. In the last section a new conceptual AM framework for handling text derived from social media is proposed and eventually, future direction for the prosperous field of AM are provided.

\section{Argumentation Theory and Models}\label{sec:ArgumentationModels}

Although argumentation mining as a term was first introduced in 2009 by Palau and Moens \cite{Palau2009ArgumentationMining}, the act of argumentation and its effects are studied since the 4th century BC \cite{Aristotle2006OnDiscourse}. Since then, many approaches on studying argumentation have been studied and many theories, schemes, and diagrams have been developed. 

The primary factor that has led to the creation of novel evaluation and visualization techniques for argument representation is the need for simple, but effective ways to break down, analyse and eventually better understand arguments. Argumentation can reach a high level of complexity, thus simpler forms of representation are needed. The process of argument illustration and fragmentation is a fundamental concept in AM, where the arguments are inspected, evaluated and eventually expressed in a binary format, capable to be interpreted from different algorithms. In the next subsection, significant argumentation theories are presented, followed by the first research works that realize the connection of argumentation with AI.

\subsection{Logical Schemes / Diagrams} \label{subesec:LogicalSchemes}

Argumentation holds its roots in rhetoric and philosophy thus argumentation diagrams have been developed as aiding tools for the analysis of arguments in well-structured documents and apart from its original role of teaching the reasoning process without falling into logical fallacies, Reed et al. \cite{REED2007ArgumentIntelligence} have also stressed their capabilities as analytical tools for meta-philosophy purposes. Diagram techniques started as practical tool for teaching logic and then developed in a more refined method that is used as a concept idea for the modeling of an argument. Logical diagrams have boosted the field of informal logic, as they offer a tool for analysing and evaluating arguments used in everyday life in a much more pragmatic environment comparing to formal logic. A detailed review on argumentation diagrams and their connection with other fields such as formal and informal logic, law, and artificial intelligence is presented in \cite{REED2007ArgumentIntelligence}. As the goal of this review paper is neither the connection of argument diagrams to modern AM techniques as in \cite{Peldszus2013FromTexts}, nor the introduction of a classification system for argumentation schemes \cite{Walton2016ASchemes} in this subsection a synopsis of the five more influential diagrams is presented and they are evaluated based on their suitability for the tasks of AM in noisy text. It has to be stressed that the AM diagrams have not been specifically designed to serve the modern construction of arguments as expressed in social media, as the tasks of detection, classification and evaluation of argumentative content in noisy text require more flexible schemes, such the one proposed in the subsection \ref{subsec:ProposedConceptualArchitecture}.

\begin{figure}[t]
\includegraphics[width=11cm, height=5cm]{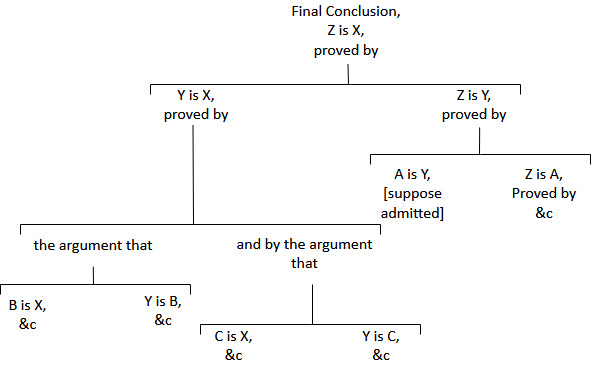}
\centering
\caption{Whately's diagram \cite{Whately1857ElementsLogic.} for analysing arguments based on backward reasoning. The final conclusion is represented as the root of the tree and its assertions are represented as leaves and the depth of the tree is proportionate to the complexity of the argument. }
\label{fig:WhatelyScheme}
\end{figure}

One of the first uses of diagrams in argumentation took place in 1857 by Whately \cite{Whately1857ElementsLogic.}, opposing to the enumeration of technical rules, in an effort of simplifying the teaching method of argumentation in his era. Whately’s diagram theory is based on the concept of identifying the concluding assertion and tracing the reasoning backwards, grounding the original assertion and eventually forming a tree with assertions and proofs. In Figure \ref{fig:WhatelyScheme}, the conclusion is represented as the root of the tree, and the assertions are located underneath. In the classical example of Socrates syllogism (Socrates is a man, all men are mortal, therefore Socrates is mortal), the conclusion (Socrates is mortal) would be the root of the tree and the two premises (p1- Socrates is a man, p2 – all men are mortal) would be the two leaves of the tree. The complexity of the reasoning process and the depth of the tree are proportionate, and could lead to complicated process that requires both well-structured arguments and experienced annotators. 

\begin{figure}[t]
\includegraphics[width=10cm, height=5cm]{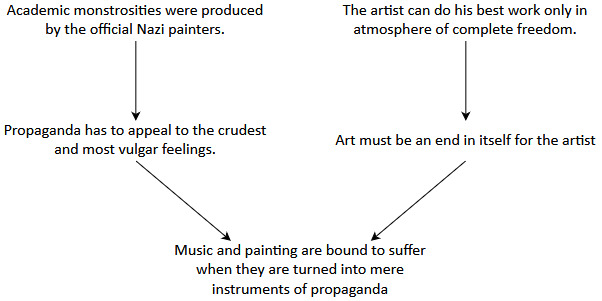}
\centering
\caption{Beardsley's convergent argument scheme \cite{Beardsley1950PracticalLogic} provided with an example. The serial premises eventually lead (converge) to the final conclusion. It should be noted that the links between the premises are not evaluated. }
\label{fig:BeardsleyArgumentSchemeWhiteBackground}
\end{figure}

Another significant work in the field of Argumentation is conducted by Beardsley in \cite{Beardsley1950PracticalLogic} introducing a designing principal that is applied until today; representation of the argument’s distinctive statements as linked nodes. As the nodes can be connected with each other with different ways, he created three basic classes 1) convergent, 2) divergent and 3) serial arguments.  In Figure \ref{fig:BeardsleyArgumentSchemeWhiteBackground}, a logic flow is illustrated depicting serial linking between premises, leading to a convergence for the final argument. In the example of the convergent argument, different premises eventually contribute towards the establishment of a reliable and robust argument, supported with enough backing.

The only flaw in Beardsley's theory is the lack of support between the statements in the nodes. Thus, the statements which form the argument are considered flawless and there are not subjects of support, debate or evaluation, hence it cannot be applied in ambiguous, implicit or imperfect arguments.

In 1958, Toulmin \cite{Toulmin1958TheArgument} suggested one very influential scheme until today, examining the role that different utterances might have in the persuasive perspective of the argument. In Toulmin's model, six functional roles were suggested, namely datum, claim, warrant, backing, qualifiers, and rebuttal.

\begin{figure}[t]
\includegraphics[width=10cm, height=5cm]{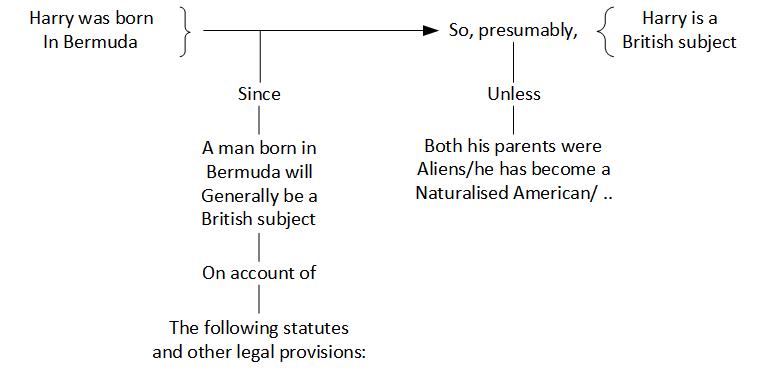}
\centering
\caption{Toulmin's proposed scheme provided with an example \cite{Toulmin1958TheArgument}. Based on its detailed structure, an argument can be assessed through the review of its distinctive components.}
\label{fig:ToulminExample}
\end{figure}

In Figure \ref{fig:ToulminExample} an example of legal nature is expressed through the Toulmin’s model. The warrant in the depicted example (\textit{“A manborn in Bermuda will generally be a British subject”}) supports adequately the initial datum (\textit{“Harry was born in Bermuda”}) as it includes a real strong backing using a legal framework (\textit{“The following statues and other legal provisions”}). The above distinctive components are enhanced through the defy of a possible counter-argument (\textit{“Both his parents were aliens..”}) and eventually lead with certainty expressed through the qualifier (\textit{So, presumably}) to the conclusion (\textit{“Harry is a British subject”}). 

The novelty on his approach lies on the fact that it requires the assignment of a predefined characterization for the cognitive connection between the different components of the argument. Through his proposed scheme, Toulmin managed to handle enthymematic relations by defining different aspects of a syllogism and connecting the inference with the warrant. The proposed scheme it is widely used in AM and a modification has even been applied in web-generated data \cite{Habernal2017ArgumentationDiscourse}, however the small IAA in some topics indicate the difficulty of applying a such complex model in heterogeneous text.

Another argumentation theory that still has a strong influence until today as recent research works \cite{Kuribayashi2018TowardsIdentification,Peldszus2015JointMining} adopt its scheme in a data driven approach for AM tasks, is developed by Freeman \cite{Freeman1991DialecticsStructure}. Freeman’s theory could be characterized as an upgrade of the Beardley’s theory, as it uses the scheme of inductive/deductive reasoning and enhances it with the concept of modality, which indicates the strength of induced conclusion by the premises. It could be said that the term of modality is an adjustment of the Toulmin's qualifier, but with a focus on the evaluation of the argument. Both the concepts of reasoning flow and argument strength have the potential to enhance the AM tasks as they offer new unprecedented tasks that have not been tested, as only sentiment flows in AM have been researched \cite{Wachsmuth2015SentimentArgumentation} until now.

\begin{figure}[t]
\includegraphics[width=10cm, height=6cm]{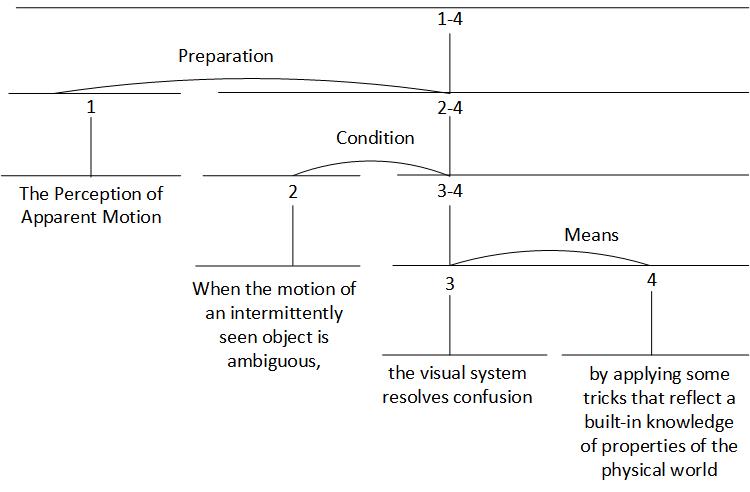}
\centering
\caption{Rhetorical Structure Theory scheme provided with an example \cite{Mann1984DiscourseGeneration}. The example that is used is the theorem of perception of apparent motion (initial nucleus), which is justified by a set of premises, and each premise is analysed consequently based on the nucleus-satellite model.}
\label{fig:RSTSchemeExample}
\end{figure}

The introduction of a prominent text-organization theory took place in the 1980's by Mann \cite{Mann1984DiscourseGeneration}, aiming at the organization of the text into different regions. Each region has a central part (nucleus) that is essential for the comprehension of the text, and a number of satellites containing additional information about the nucleus. The nucleus and the satellite are correlated with each other through different relations (circumstance, elaboration, evidence, etc.) which can be changed, manipulated, added or subdivided depending on the topic and the task at hand. The nucleus-satellite distinction is applied recursively until every entity of the discourse is a part of a RST relation and eventually a tree-structure hierarchy is created as depicted in Figure \ref{fig:RSTSchemeExample}, where the theorem of the perception of apparent motion (initial nucleus) is supported by a set premises, where each premise sequentially is expressed through the model of nucleus-satellite expressing a specific relation (preparation, condition, means). The application of RST has improved the performance of sentiment polarity classification when enhanced with argumentation \cite{Carstens2017UsingProblems}, but the authors state the need for further in-depth evaluation for its impact.

\subsection{Early Argumentation Theory in  AI}

Argumentation diagrams have boosted the entire field of computational linguistics, but as they are theoretical models, they do not have the capacity to reform the entire argumentation field. In the 1990's, the first research connecting argumentation with the area of Artificial Intelligence (AI) was conducted. These early attempts on connecting arguments with AI created a new field under the name Computational Argumentation, the field which has formed AM in a great extent. Computational Argumentation is used extensively in domains where reasoning process is sophisticated, such as law or medicine, and therefore traditional methods like formal logic and classical decision theory cannot be applied.

One of the first researches studying the relationship between argumentation, cognitive psychology and AI was conducted by Pollock \cite{Pollock1987DefeasibleReasoning} describing the connection of defeasible reasoning in philosophy and nonmonotonic reasoning in AI through the definition of a set of rules. An important step ahead was made by Dung  \cite{Dung1995OnGames}, who researched the relation between argumentation and Logic Programming, focusing on the modelling of the fundamental mechanisms humans use in argumentation and expressing them with a number of rules and definitions capable to be interpreted by AI algorithms.

Similar to Dung's method, Krause et al. \cite{Krause1995AUncertainty} developed the logic of argumentation, an approach for defining reasoning in cases of uncertainty. By defining rules, definitions and propositions they have managed to create a theoretical system capable to estimate the strength of an argument based on the distinctive axioms used for its construction. The degree of justification was also researched by Pollock \cite{Pollock2001DefeasibleJustification}, focusing on the different degrees of justification the distinctive components of an argument can provide when they are \textit{"summed"}.

Another work that indicates the relationship between argumentation and logic programming was held by Parsons et al. \cite{Parsons1996NeogotiationReport}, where a framework was developed capable to interpret a broad range of negotiation in a multi-agent system. The distinctive arguments of the agents are considered as logical steps towards to an acceptable compromise, not necessarily towards the optimum proposal. A more detailed review on the automated negotiation is held in \cite{Jennings2001AutomatedChallenges}  presenting a more generic framework and three different approaches (game theoretic models, heuristics, argumentation-based) for the negotiation process. The development of a dialectical argumentation scheme was introduced in \cite{GRASSO2000DialecticalNutrition}, where the authors implemented a fully-functional agent capable to maintain the natural flow on the debate by arguing on a topic and in the same time responding to obligations related to the discourse.

Although many of the aforementioned frameworks are designed to cover argumentation in its full generality, any system with pre-defined rules is not suggested for use in the open and constantly changing environment of social media, as it does not have the ability to learn and adopt. In Table \ref{schemes_comparison}, a synopsis of the theories and schemes that have been presented in this section are illustrated and evaluated based on a number of criteria focusing on their suitability on AM in social media, but the table could could also be used as a point of reference for tasks in the wider area of NLP.

The first criteria of a theory's evaluation is the explicit expression of any kind of relations between the distinctive components of the argument, a property that can facilitate advanced NLP tasks such as relations identification, evidence detection, and facts recognition. In the theories that are based on logic programming, the relations are expressed as rules, whereas in Toulmin's theory is expressed through the qualifier and in Beardley's as modality. The second criteria in the comparison table is the level of complexity, which is determined based on the number of components and relations each theory involves. Both the in-depth analysis and the construction of the reverse tree in Whately's theory, and the definition of six different components in Toulmin's diagram present as expected high levels of complexity. Once again logic programming theories can be grouped together presenting high complexity as they define multiple rules and axioms in order to cover a wide range of cases. A special note should be given in RST, as it is a flexible open scheme, where components and relations can be defined and modified based on each case study. The first assessment of the theories/schemes take place based on their suitability to be applied in NLP tasks in noisy environment. The complexity level of a theory is inversely proportional to its suitability when applied on noisy environment as complicated reasoning processes cannot be deployed in text lacking grammatical and structural rules. Finally, the last column of the table illustrates the applicability level of the theories for AM tasks independently from the source of the content, where three schemes stand out (Beardsley, Toulmin, RST), each one for different reasons. Beardsley's theory is straightforward enough to be applied easily in a wide range of goals, Toulmin's detailed scheme is the option for in-depth analysis and RST can be easily modified in order to cover the needs and the requirements of each case study.

\begin{table}[]
\centering
{%
\begin{tabular}{|l|l|l|l|l|}
\hline
Author(s) & {\begin{tabular}[c]{@{}l@{}}Link\\ Relation\end{tabular}} & {\begin{tabular}[c]{@{}l@{}}Complexity\\ Level\end{tabular}} & {\begin{tabular}[c]{@{}l@{}}Application in\\ noisy data\end{tabular}}    & {\begin{tabular}[c]{@{}l@{}}Application \\ in AM \end{tabular}} \\ \hline
Whately         & No & High & Low & Medium  \\ \hline
Beardsley       & No & Low & Medium & High  \\ \hline
Toulmin         & Yes & High & Medium-Low & High  \\ \hline
Freeman         & Yes & Medium & Medium & Medium \\ \hline
RST             & Yes & Open & High & High \\ \hline
Pollock         & Yes & High & Low & Medium  \\ \hline
Dung            & Yes & High & Low & Medium  \\ \hline
Krause          & Yes & High & Low  & Medium  \\ \hline
Parsons         & Yes & High & Low & Medium  \\ \hline
New Rhetoric    & Yes & High & Low & Medium  \\ \hline
\end{tabular}%
}
\caption{A synopsis of logical schemes and computational theories based on the degree they can meet the needs in a modern NLP environment. Link relation - if the connection distinctive components are explicitly evaluated, Complexity level - is assessed based on the number of components each theory includes.   }
\label{schemes_comparison}
\end{table}

Computational argumentation contributed significantly in the creation of the AM field, but the strict rules that used to be posed seem outdated. Although the foundations of logic programming are based on knowledge- and logic-based solutions, the recent advances in ML have been used extensively in data-driven approaches \cite{Stab2018Cross-topicNetworks, Addawood2017StanceCase}, but they seem to have reached the upper limit of their capabilities.  Rule-based machine learning could be considered it as a data-driven approach of inductive logic programming that combines background knowledge and ability to learn based on human readable theories. Another approach combining a data-driven solution in unison with argumentation structure took place in \cite{Carstens2017UsingProblems} where the probabilistic classifiers have been improved with the incorporation of an argument database.

\section{The status of AM in Social Media}
\label{sec:AMSocialMedia}

AM is considered a complicated task because it challenges a series of closely interrelated distinctive tasks that come under this general term. The classification of arguments consists from a series of steps and it is considered the final output of the pipeline, however, other distinctive parts of the pipeline, such as argument detection, could form the main research question. Especially in the world of social media (and web-generated context in general) tasks that are focused on the reliability and the strength of the arguments seem to gain the interest of the research community. In the rapidly changing environment of social media, the presence or not of argumentative features can assist scholars in the tasks of rumour spreading, fake news detection and source identification.  

A recent survey \cite{Lippi2016ArgumentationMining} suggests a general-purpose pipeline for the task of AM with two main components, argument component detection and argument structure prediction. The proposed architecture can be generalized in every source of raw text without significant problems, however if it has to be applied in AM tasks from text derived from social media would face some obstacles, as either important generalizations would take place either some tasks would be ignored. In our effort to discover better the connection between AM and social media we describe the basic tasks that take place in an AM pipeline and compare the results gained from different research methods.  In \ref{subsec:ArgumentDetection} the task of argument detection is presented, followed by the tasks of relations identification in \ref{subsec:RelationsIdentification} and stance detection in \ref{subsec:StanceDetection}. In the last sub-section, different tasks that are related with the concept of argument reliability in social media are presented.

\subsection{Argument Detection}
\label{subsec:ArgumentDetection}


\begin{table}[]
\centering
\resizebox{\textwidth}{!}{%
\begin{tabular}{|l|l|l|l|l|l|l|l|}
\hline
\multirow{2}{*}{Authors} & \multirow{2}{*}{Task(s)} & \multirow{2}{*}{Size} & \multirow{2}{*}{Topic} & \multirow{2}{*}{\#Annot.} & \multicolumn{2}{l|}{\multirow{2}{*}{IAA}} & \multirow{2}{*}{\#Classes} \\
 &  &  &  &  & \multicolumn{2}{l|}{} &  \\ \hline
 
\multirow{2}{*}{Addawood and Bashir \cite{Addawood2016WhatMedia}} & \begin{tabular}[c]{@{}l@{}}Argument\\ Detection\end{tabular} & \multirow{2}{*}{3000  tweets} & \multirow{2}{*}{\begin{tabular}[c]{@{}l@{}}Apple/FBI\\ encryption debate\end{tabular}} & \multirow{2}{*}{2} & \multicolumn{2}{l|}{0.67 Ck} & 2 \\ \cline{2-2} \cline{6-8} 
 & \begin{tabular}[c]{@{}l@{}}Evidence\\ Categorization\end{tabular} &  &  &  & \multicolumn{2}{l|}{0.79 Ck} & 6 \\ \hline
 
\multirow{2}{*}{Bosc et al. \cite{Bosc2016TweetiesMedia.}} & \begin{tabular}[c]{@{}l@{}}Argument\\ Detection\end{tabular} & \multirow{2}{*}{4000 tweets} & \multirow{2}{*}{5 debate topics} & \multirow{2}{*}{3} & \multicolumn{2}{l|}{0.81 Ka} & 2 \\ \cline{2-2} \cline{6-8} 
 & \begin{tabular}[c]{@{}l@{}}Relations\\ Identification\end{tabular} &  &  &  & \multicolumn{2}{l|}{0.67 Ka} & 3 \\ \hline
 
Deturck et al. \cite{Deturck2018ERTIMMC2:Retrieval} & \begin{tabular}[c]{@{}l@{}}Argument\\ Detection\end{tabular} & 70 M tweets & Festival evaluation & - & \multicolumn{2}{l|}{-} & - \\ \hline

\multirow{3}{*}{Dusmanu et al. \cite{Dusmanu2017ArgumentSources}} & \begin{tabular}[c]{@{}l@{}}Argument\\ Detection\end{tabular} & \multirow{3}{*}{1887} & \multirow{3}{*}{Grexit / Brexit} & \multirow{3}{*}{2} & \multicolumn{2}{l|}{0.77 Ck} & 2 \\ \cline{2-2} \cline{6-8} 
 & \begin{tabular}[c]{@{}l@{}}Factual vs\\ opinion\end{tabular} &  &  &  & \multicolumn{2}{l|}{0.73 Ck} & 2 \\ \cline{2-2} \cline{6-8} 
 & \begin{tabular}[c]{@{}l@{}}Source\\ Identification\end{tabular} &  &  &  & \multicolumn{2}{l|}{0.84 Dice} & 2 \\ \hline
 
Sendi and Latiri \cite{Sendi2018OpinionModeling.} & \begin{tabular}[c]{@{}l@{}}Argument \\ Detection\end{tabular} & 70 M tweets & Festival evaluation & - & \multicolumn{2}{l|}{-} & - \\ \hline

Dufour et al. \cite{Dufour2018LIACLEFNetwork.} & \begin{tabular}[c]{@{}l@{}}Argument\\ Detection\end{tabular} & 70 M tweets & Festival evaluation & - & \multicolumn{2}{l|}{-} & - \\ \hline

\multirow{2}{*}{Cocarascu and Toni \cite{Cocarascu2018CombiningDatasets}} & \multirow{2}{*}{\begin{tabular}[c]{@{}l@{}}Relations\\ Identification\end{tabular}} & 30 & \multirow{2}{*}{\begin{tabular}[c]{@{}l@{}}Hillary Clinton's\\ FBI investigation\end{tabular}} & 3 & \multicolumn{2}{l|}{0.11 Fk} & 3 \\ \cline{3-3} \cline{5-8} 
 &  & 840 &  & NG & \multicolumn{2}{l|}{NG} & NG \\ \hline
 
Ma et al. \cite{Ma2018ClaimTwitter} & \begin{tabular}[c]{@{}l@{}}Relations\\ Identifications\end{tabular} & 2520 & 30 debate topics & 3 & \multicolumn{2}{l|}{0.81 MV} & 2 \\ \hline

Mohammad et al. \cite{Mohammad2017StanceTweets} & \begin{tabular}[c]{@{}l@{}}Stance\\ Detection\end{tabular} & 4163 & 5 debate topics & 5-8 & \multicolumn{2}{l|}{0.82 MV} & 3 \\ \hline

Zarella and Marsh \cite{Zarrella2016MITREDetection} & \begin{tabular}[c]{@{}l@{}}Stance\\ Detection\end{tabular} & 4163 & 5 debate topics & 5-8 & \multicolumn{2}{l|}{0.82 MV} & 3 \\ \hline

Wei et al. \cite{Wei2018Multi-TargetNetwork} & \begin{tabular}[c]{@{}l@{}}Stance\\ Detection\end{tabular} & 4163 & 5 debate topics & 5-8 & \multicolumn{2}{l|}{0.82 MV} & 3 \\ \hline

Ebrahimi et al. \cite{Ebrahimi2016WeaklyBootstrapping} & \begin{tabular}[c]{@{}l@{}}Stance\\ Detection\end{tabular} & 78000+ & Donald Trump & - & \multicolumn{2}{l|}{-} & - \\ \hline

Lai et al. \cite{Lai2018StanceDebate} & \begin{tabular}[c]{@{}l@{}}Stance\\ Detection\end{tabular} & 963 triplets & \begin{tabular}[c]{@{}l@{}}Italian Constitution \\ refenderum\end{tabular} & 2-5 & \multicolumn{2}{l|}{} & 3 \\ \hline

Johnson and Goldwasser \cite{Johnson2016IdentifyingTwitter} & \begin{tabular}[c]{@{}l@{}}Stance\\ Detection\end{tabular} & 99161 & 16 debate topics & - & \multicolumn{2}{l|}{-} & - \\ \hline

\multirow{2}{*}{Addawood et al. \cite{Addawood2017StanceCase}} & \begin{tabular}[c]{@{}l@{}}Stance\\ Detection\end{tabular} & \multirow{2}{*}{3000} & \multirow{2}{*}{\begin{tabular}[c]{@{}l@{}}Apple/FBI \\ encryption debate\end{tabular}} & \multirow{2}{*}{2} & \multicolumn{2}{l|}{0.64 Ck} & 3 \\ \cline{2-2} \cline{6-8} 
 & \begin{tabular}[c]{@{}l@{}}Topic\\ Classification\end{tabular} &  &  &  & \multicolumn{2}{l|}{0.70 Ck} & 4 \\ \hline
 
\multirow{2}{*}{Konstantinovskiy et al. \cite{Konstantinovskiy2018TowardsDetection}} & \multirow{2}{*}{\begin{tabular}[c]{@{}l@{}}Claim\\ Detection\end{tabular}} & \multirow{2}{*}{4080} & \multirow{2}{*}{\begin{tabular}[c]{@{}l@{}}Various\\ political issues\end{tabular}} & \multirow{2}{*}{5} & \multicolumn{2}{l|}{0.46 Ka} & 7 \\ \cline{6-8} 
 &  &  &  &  & \multicolumn{2}{l|}{0.7 Ka} & 2 \\ \hline
\end{tabular}%
}
\caption{Ck - Cohen's kappa \cite{Carletta1996AssessingStatistic}, Ka - Krippendorff's alpha \cite{krippendorff2004MeasuringData}, Dice \cite{Dice1945MeasuresSpecies}, Fk - Fleiss kappa \cite{Fleiss1971MeasuringRaters.}. In \cite{Ma2018ClaimTwitter}, the IAA is the average of the three sub-tasks. 
Details of datasets that have been used in recent research papers, displaying the distinctive tasks in AM, the different sizes of the datsaets, the plethora of topics and the range in IAA metrics.  }
\label{tab:dataset-comparison}
\end{table}


The classification of a sentence or a series of sentences as argumentative or non-argumentative is a crucial step towards AM in social media. Argument detection is in essence a preliminary binary classification that would enable the subsequent in-depth analysis of the argument, such as persuasiveness detection, relations identification between the components of the argument or automatic evaluation of the argument.

Data from social media are a characterized as a special category of web-generated data and this becomes clear if an attempt is made on applying the Toulmin scheme in a typical tweet. In the work of Addawood and Bashir \cite{Addawood2016WhatMedia} the following tweet from their dataset is presented: \textit{RT @ItIsAMovement "Without strong encryption, you will be spied on systematically by lots of people" - Whitfield Diffie}. The above tweet cannot easily fit in the Toulmin's scheme (or any other theoretical scheme) as the fact consists the same time the conclusion of the argument, whereas the component of the backing is expressed through the quote of an expert opinion. A similar belief is also expressed in \cite{Bosc2016TweetiesMedia.} where the authors claim \textit{"we (almost) never find such a kind of complete structure of the arguments".}

Towards to the analysis of the tweet, the first step is defining the necessary background knowledge that a reader/annotator has on the discussed topic, in order the IAA not to be affected by the different level of the annotators’ knowledge. For the ease of the annotation task some general rules can be defined that can provide a useful insight for the nature of the text. When the user cites the opinion of an expert, express a feeling or irony it is very likely the tweet to be argumentative. On the other hand, if the tweet is the title of an article and simply shares a link without any comments, the tweet can be characterized as non-argumentative. An example of a non-argumentative tweet is provided in the work of Dusmanu et al. \cite{Dusmanu2017ArgumentSources} regarding Brexit: \textit{72\%  of  people  who  identified  as “English” supported   \#Brexit   (while   no   majority   among those   identifying   as   “British”) https://t.co/MuUXqncUBe}.

The rules that have been described should be consider more as guidelines, as there are many cases where a tweet cannot easily fit any of the proposed categories. In many cases a twitter user takes into account more information than simply the information contained in the tweet, as the beliefs or the status of the user who tweeted. Thus, defining a specific level of knowledge of the annotators is crucial for every use case.

Other important factors that affect the annotation process and the quality of the dataset are the requested task, the size of the dataset, the debate topic and the number of the annotators that have been used for carrying out the specific task. The above aspects can be used as metrics for measuring the quality of the collected data, as often scholars put thresholds in IAA for accepting or rejecting specific datasets. Table \ref{tab:dataset-comparison} provides a detailed review of the characteristics of the datasets that have been used in recent research papers and are analysed in the following subsections. The table includes both annotated datasets that have been used in supervised ML approaches, but also datasets that have been used in un-supervised approaches. There is a scarce of annotated datasets in AM and most importantly there is a difficulty in their re-use, as they are often annotated for very specific tasks. The recent review paper of \cite{Cabrio2018FiveAnalysis} provides a complete synopsis of the available datasets in the area of AM, without focusing on social media generated text.

Based on the results depicted in Table \ref{tab:dataset-comparison}, the IAA depends heavily from the task at hand as it is revealed in Addawood et al. \cite{Addawood2017StanceCase} and in Addawood and Bashir \cite{Addawood2016WhatMedia}. On those two research papers, the same annotators, in the same dataset present higher IAA for the tasks of Evidence/Topic Classification in comparison to the tasks of Argument/Stance Detection, although the later offer more possible classes from the former. In the case of limiting the possible classes for the same task, as in \cite{Konstantinovskiy2018TowardsDetection}, the IAA increases as it is expected, but the task ceases to provide a detailed analysis. The last point that should be stressed considering the IAA is the different metrics that are used in the different studies, as  the number of annotators affects the available options.

The task of argument detection as a necessary preliminary step takes place in the work of Addawood and Bashir \cite{Addawood2016WhatMedia} and also in Dusmanu et al. \cite{Dusmanu2017ArgumentSources}, where the proposed pipelines end up in the recognition of evidence type and in the source identification respectively. In both works the detection of argumentative tweets is necessary, as the non-argumentative tweets consist respectively the 42.3\% and the 29.3\% of the constructed datasets. Both papers adopt a supervised approach where a manual annotation is required, succeeding a substantial inter-annotation agreement (IAA) in terms of Cohen's kappa equal to 0.67 and 0.77, respectively. A similar architecture was also adopted in \cite{Bosc2016TweetiesMedia.} ending up in the construction of argumentation graphs. The DART dataset \cite{Bosc2016DART:Scholar} was used as input of the pipeline, containing 2702 argumentative tweets and 1181 non-argumentative tweets. In contrast to \cite{Addawood2016WhatMedia, Dusmanu2017ArgumentSources}, Bosc et al. \cite{Bosc2016TweetiesMedia.} measured the IAA in terms of Krippendorff’s alpha, resulting to the satisfactory a=0.81.  

On the other hand, a semi-supervised approach was followed in the papers presented in CLEF 2018 conference for the task of Multilingual Cultural Mining and Retrieval. In the work of Deturck et al. \cite{Deturck2018ERTIMMC2:Retrieval} the assumption that an argumentative text is structured to effectively combine arguments and opinions takes place, thus argumentation is measured through structuration. A similar approach was followed in Sendi and Latiri \cite{Sendi2018OpinionModeling.}, where argumentative tweets are defined as the sum of three separate tasks: information retrieval, topic modeling and sentiment score. The work from Dufour et al. \cite{Dufour2018LIACLEFNetwork.} follows a distance supervision approach through the detection of five pre-defined features: emotion words, emoticons, particular punctuation signs, personal pronouns and hashtags.

\begin{table}[]
\resizebox{\textwidth}{!}{%
\begin{tabular}{|l|l|l|l|l|l|l|}
\hline
                     & Basic (Lexical)                                                        & Semantic                                                                                                      & Sentiment                                                                  & Subjectivity                                                            & Twitter    & Other                                                                                                                                                                                     \\ \hline
Addawood and Bashir \cite{Addawood2016WhatMedia}   &  \begin{tabular}[c]{@{}l@{}}n-gram, \\ length, \\ question, \\ exclamation \end{tabular}                                                                 & \begin{tabular}[c]{@{}l@{}}PoS, \\ LIWC summary, \\ variables \end{tabular}                             &  \begin{tabular}[c]{@{}l@{}}LIWC, \\ sentiment, \\ lexicon \end{tabular}      & \begin{tabular}[c]{@{}l@{}}Clue   lexicon  \end{tabular} & \begin{tabular}[c]{@{}l@{}}followers, \\ friends, \\ user activity, \\ URL+title, \\ hashtags, \\ verified account, \\ mentions\end{tabular} & \begin{tabular}[c]{@{}l@{}}Psycometric, \\ LIWC \end{tabular}  \\ \hline

Bosc et al. \cite{Bosc2016TweetiesMedia.}   &  \begin{tabular}[c]{@{}l@{}}n-gram, \\ punctuation, \\ tokens, \\ capitalization \end{tabular}                                                                 & \begin{tabular}[c]{@{}l@{}}PoS \end{tabular}                             &      &    & smileys  &    \\ \hline

Deturck et al. \cite{Deturck2018ERTIMMC2:Retrieval}    & \begin{tabular}[c]{@{}l@{}}tokens, \\  lemmas\end{tabular}           & PoS                                                                                                   & \begin{tabular}[c]{@{}l@{}}TextBlob  \\ features  \end{tabular}   & \begin{tabular}[c]{@{}l@{}}TextBlob \\ features \end{tabular}        &          & \begin{tabular}[c]{@{}l@{}}word \\ embeddings \end{tabular}                                                                                                                                                                                      \\ \hline

Dusmanu et al. \cite{Dusmanu2017ArgumentSources}    & \begin{tabular}[c]{@{}l@{}}n-gram \end{tabular}      & \begin{tabular}[c]{@{}l@{}}Syntactic tree, \\ parse trees, \\ dependency relations \\ WordNet synset \end{tabular} & AlchemyAPI   &                                                                         & \begin{tabular}[c]{@{}l@{}}Punctuations, \\ emoticons\end{tabular}     &                                                                                                                         \\ \hline

Sendi and Latiri \cite{Sendi2018OpinionModeling.} & \begin{tabular}[c]{@{}l@{}}IR, \\ TM\end{tabular} &                                                                                                                              & \begin{tabular}[c]{@{}l@{}}NRC \\  emotion lexicon \end{tabular}          &                                                                         &         &                                                                                                                                                                                          \\ \hline

Dufour et al. \cite{Dufour2018LIACLEFNetwork.}    & punctuation,                                                           & personal pronouns                                                                                                              & Emotion words                                                              &                                                                         & \begin{tabular}[c]{@{}l@{}}Emoticons, \\ hashtags\end{tabular}           &                                                                                                                      \\ \hline

Cocarascu and Toni \cite{Cocarascu2018CombiningDatasets} & n-grams    &       &           &           & 
&   \begin{tabular}[c]{@{}l@{}} word embeddeings \\ argumentativeness \end{tabular} \\ \hline

Ma et al. \cite{Ma2018ClaimTwitter}    & \begin{tabular}[c]{@{}l@{}} n-gram, \\ Relevance \\ feature, \\ Okapi BM25 \end{tabular}      & \begin{tabular}[c]{@{}l@{}} Topic-Independent \\ Claim-Related \\ Lexicon, \\ Topic-Dependent \\ Claim-Oriented\\ Lexicon \end{tabular} &  &  \begin{tabular}[c]{@{}l@{}} Controversy \\lexicon   \end{tabular}     & \begin{tabular}[c]{@{}l@{}}URLs, \\ retweet \\ reply \end{tabular}     &  \begin{tabular}[c]{@{}l@{}}  TwitStan   \\ WikiClaim     \\ TwitArgument   \end{tabular} \\
\hline

Mohammad et al. \cite{Mohammad2017StanceTweets} & n-grams & PoS & \begin{tabular}[c]{@{}l@{}}1)NRC Emotion\\ Lexicon,\\ 2)Hu and Liu \\Lexicon\end{tabular} &   \begin{tabular}[c]{@{}l@{}} MPQA \\ Subjectivity \\ Lexicon \end{tabular}  & hashtag & \begin{tabular}[c]{@{}l@{}}Target of interest,\\ encodings\end{tabular} \\ \hline

Zarella and Marsh \cite{Zarrella2016MITREDetection}  &     &       &           &           &           & word embeddings    \\ \hline

Wei et al. \cite{Wei2018Multi-TargetNetwork} &  &  &  &  &  & \begin{tabular}[c]{@{}l@{}} word embeddings, \\ target content \end{tabular} \\ \hline

Ebrahimi et al. \cite{Ebrahimi2016WeaklyBootstrapping} & \begin{tabular}[c]{@{}l@{}}n-grams,\\ linguistic \\patterns\end{tabular} &  & \begin{tabular}[c]{@{}l@{}}1) LIWC2007\\ 2) VADER\end{tabular} &  &  &  \\ \hline

Lai et al. \cite{Lai2018StanceDebate}   &     &           &           &           &    \begin{tabular}[c]{@{}l@{}} social media \\network communities, \\Hashtags,\\ Mentions, Replies \end{tabular}        & \\ \hline

Johnson and Goldwasser \cite{Johnson2016IdentifyingTwitter} & \begin{tabular}[c]{@{}l@{}} Words\\ frequency \end{tabular}  & \begin{tabular}[c]{@{}l@{}}keyword-based\\ heuristic \end{tabular} & OpinionFinder 2.0 &  & Temporal activity &  \\ \hline

Addawood et al. \cite{Addawood2017StanceCase} & n-grams & PoS & \begin{tabular}[c]{@{}l@{}} Subjectivity\\ lexicon \end{tabular} & \begin{tabular}[c]{@{}l@{}} MPQA \\ Subjectivity \\ Lexicon \end{tabular}  & \begin{tabular}[c]{@{}l@{}}RT, title, mention, \\ verified account, url, \\ followers, following, \\ posts, hashtag\end{tabular} & \begin{tabular}[c]{@{}l@{}}Argumentativeness,\\ source type\end{tabular} \\ \hline

Konstantinovskiy et al. \cite{Konstantinovskiy2018TowardsDetection}     &       & PoS, NER, tf-idf       &       &       &   & \begin{tabular}[c]{@{}l@{}}sentence \\ embeddings \end{tabular} \\   \hline

\end{tabular}%
}
\caption{Features used for AM tasks in social media. IR = Information Retrieval applied as lexicon-based queries, PoS = Part of Speech, TM = Topic Modeling applied with the technique of Latent Dirichlet Allocation (LDA), LIWC = Linguistic Enquiry and Word Count \cite{Pennebaker1996CognitiveDisclosure, Pennebaker1997WritingProcess}
}
\label{FeaturesTable}
\end{table}

The features used in the aforementioned research, but also for those that they are going to be presented in the next sections, are provided in Table \ref{FeaturesTable}, where are summarized in 5 different categories. Lexical features are the attributes that are used most frequently in the wider spectrum of the NLP and they are strongly correlated with the different applications of n-grams, whereas as semantic features we define the characteristics of the language that can provide a deeper insight of the data. Sentiment features are those who can trigger emotions and usually they are detected with the use of specific lexicons or libraries and the subjectivity features often indicate an opinionated and therefore an argumentative tweet. The twitter-specified features are offered as metadata through the twitter API and concern the specific characteristics a tweet contains, and in the last column we have collected the features that they cannot be grouped under any of the previous categories. Apart from the semantic and sentiment features, LIWC offer statistics that include personal concerns, core drives and needs, which are summarized under the psychometric category. In the work of Deturck et al. \cite{Deturck2018ERTIMMC2:Retrieval} the use of word embeddings takes place aiming at a diversity filtering in order the most argumentative tweets to be discovered.   

It has to be mentioned that the classification of features is different in \cite{Addawood2016WhatMedia} (the only work providing a detailed table of features), classifying emotional tone and subjectivity score under the linguistics features. Furthermore, we include the tasks of information retrieval and topic modeling as described in \cite{Sendi2018OpinionModeling.} as lexical feature, provided that they do not discover any semantic purpose. Those adjustments took place in order to provide a useful taxonomy and a beneficial comparison, but it should be noted that different categorizations can take place.

Regarding the results of the classification, those can be extracted from the works that have followed a supervised approach and they are presented in Table \ref{ArgumentDetectionResults}. In the first column the names of the authors are depicted, in the second and the third columns the algorithms and the features that have been used are displayed respectively. The last three columns depict the metrics that can evaluate the performance of the algorithms. Concerning the feature selection and their impact for the classification task, the use of all the possible features performs better in each case. The selection of the classification algorithm does not seem to affect significantly the performance of the task in \cite{Dusmanu2017ArgumentSources}, whereas in \cite{Addawood2016WhatMedia} the use of SVM surpasses the alternative algorithms. For the task of argument detection, the best results are achieved in \cite{Addawood2017StanceCase} with 0.89 F1, whereas \cite{Dusmanu2017ArgumentSources} and \cite{Bosc2016TweetiesMedia.} achieve 0.78 F1 and 0.67 F1 respectively. The results of the semi-supervised approaches are measured in terms of ranking quality either using the Normalized Discounted Cumulative Gain (NDCG) either simply presenting qualitative results. 

\begin{table}[]
\resizebox{\textwidth}{!}{%
\centering
\begin{tabular}{|l|l|l|l|l|l|}
\hline
          Authors          &  Algorithm & Features                & Precision & Recall & F1   \\ \hline

Addawood and Bashir \cite{Addawood2016WhatMedia} & DT  &  n-gram     & 0.72      & 0.69   & 0.66 \\ \cline{2-6} 
                          & SVM & n-gram    & 0.81        & 0.78   & 0.77 \\ \cline{2-6} 
                          & NB & n-gram     & 0.70      & 0.67   & 0.64 \\ \cline{2-6} 
                          & DT & all features & 0.87      & 0.87   & 0.87 \\ \cline{2-6} 
                          & SVM & all features & 0.89      & 0.89   & 0.89 \\ \cline{2-6} 
                          & NB & all features  & 0.79      & 0.79   & 0.85 \\ \hline

Bosc et al. \cite{Bosc2016TweetiesMedia.}  & LR & lexical  &   -    & -    & 0.64 \\ \cline{2-6} 
                          & LR   & \begin{tabular}[c]{@{}l@{}}lexical  + \\ semantic\end{tabular}  & -     & -     & 0.66 \\ \cline{2-6} 
                          & LR   &  all features  &  - & -    & 0.67 \\ \hline

Dusmanu et al. \cite{Dusmanu2017ArgumentSources}  & RF & n-gram             & 0.76      & 0.69   & 0.71 \\ \cline{2-6} 
                          & LR   & n-gram       & 0.76      & 0.71   & 0.73 \\ \cline{2-6} 
                          & LR   &  all features   & 0.80      & 0.77   & 0.78 \\ \hline

\end{tabular}%
}
\caption{The results of the supervised ML algorithms for the task of argument detecton. Un-supervised methods are not included.
RF = Round Forest, LR = Logistic Regression, DT = Decision Tree, SVM = Support Vector Machine, NB = Naive Bayes. Rounding took place in order the results to be displayed in the same scale.}
\label{ArgumentDetectionResults}
\end{table}

\subsection{Relations Identification}
\label{subsec:RelationsIdentification}

The choice and the customization of the theoretical argumentation model that will be adopted in any research project affects the individual tasks that will be raised. Especially the task of relations identification, or argument structure prediction as expressed in \cite{Lippi2016ArgumentationMining}, is the most susceptible part of the AM pipeline to potential changes in the adopted model. The task of annotating relations between parts of text requires the adoption of a holistic approach, capable to identify connections with both preceding and succeeding components of the selected model, the relations between the entities of the network and eventually offer a better understanding of the argument.

Both Toulmin's \cite{Toulmin1958TheArgument} and Freeman's \cite{Freeman1991DialecticsStructure} theories, two of the most influential theories in the wider field of logic and argumentation, explicitly define relations between the components of the arguments. In data derived from social media, argument's component identification is a challenging task as both their size and their chaotic nature do not allow strict rules and principles to be applied. As a consequence of this situation, the task of relations identification should be redefined and include both micro and macro analysis. The micro-analysis evaluates the quality and the completeness of the argument, whereas the macro-analysis expresses the relation of an argument either towards a known topic or towards an argument previously expressed. In social media, network analysis algorithms can significantly boost macro-analysis tasks, as the introduction of network-based features reveal underlying relations between the users and eventually improve the prediction model \cite{Lai2018StanceDebate}.

The possible outcomes of a macro-analysis in social media text are limited to support/attack/neither relations indicating in a great extent the outcome of the stance detection. On the other hand, micro-analysis is related more with AM and other reliability-related tasks, as it evaluates the integrity and the cohesion of the argument. The arguments extracted from online resources are not characterized as high-quality data, as often complicated reasoning process should take place in order the argument to be understood. Arguments with missing premises are called enthymemes \cite{Rajendran2016ContextualReviews, Rajendran2018IsDataset} and take place often in informal discourse, creating a challenge for the approach that should be followed; discard the argument or try to fill the missing premise.

The need for both micro and macro analysis for the task of relations identification, in combination with the low-quality data from social media creates the need for the establishment of simple, but effective rules and standards. Towards the need for providing a straightforward definition able to capture both micro and macro analysis, we define three entities (argument, topic, completeness) able to capture the complicated nature of the task. We convert the problem to a mathematical expression in a triple, where the task of relations' identifications is split in two parts, where the first part connects the argument with a specific problem (favor/against/neither) and the second part evaluates the structure of the argument. Eventually, expressing the relations identification as a triple we have:
\[ (a_{\text{ij}}, t_i, c_j) \]
    
where the expressed argument \textit{a\textsubscript{ij}} is open to a macro-analysis considering a topic \textit{t\textsubscript{i}} and a micro-analysis for its completeness \textit{c\textsubscript{j}}. 

Only few researchers have explored the task of relations identification in text derived from social media, because of the chaotic nature of social media and the wide presence of vague claims. In subsection 3.1 the first step (argument detection) in the proposed pipeline of Bosc et al. \cite{Bosc2016DART:Scholar} was presented, which is followed by the prediction of attack/support relations between tweets and arguments. Their adopted approach is similar to textual entailment, thus the Excitement Open Platform (EOP) and the Recognizing Textual Entailment (RTE) framework were used. A second method was also applied implementing a neural sequence classifier, however none of the methods presented satisfactory results. In fact, the detection of support-relation achieved 0.20 F1 and the attack-relation 0.16 F1 using neural model, and with use of EOP+RTE the support-relation achieved 0.17 F1 and the attack-relation 0.0 F1. Apart from the low score in the automatic detection of relations, even the IAA was significantly lower a=0.67 for the specific task, comparing to IAA for the task of argument detection which reached a=0.81.

A new method for extracting argumentative relations of attack, support or neither is presented in \cite{Cocarascu2018CombiningDatasets} based on the Relation-based Argumentation Mining (RbAM) model. The proposed model was tested in the dataset of Carstens and Toni \cite{Carstens2017UsingProblems} and afterwards it was applied on the task of relations prediction between tweets and news headlines on two different datasets \cite{Guo2013LinkingMedia} \cite{Tan2017SpotTwitter.}. Apart from different implementations of neural networks, the impact of trained and non-trained was also evaluated demonstrating the supremacy of the trained embeddings. Apart from the use of word embeddings and the argumentativeness features extracted from RbAM, the authors do not describe the rest of the features, instead they simply use the term standard features, thus safe conclusions for the use of features cannot be drawn.

Broadening the limits of relations prediction task, Ma et al. \cite{Ma2018ClaimTwitter} introduced a 3-step framework including both micro and macro-analysis of the argument, in contrast to \cite{Bosc2016DART:Scholar} and \cite{Cocarascu2018CombiningDatasets} where only a macro analysis took place. Besides the attack/support relation between a tweet and a topic, the authors also examined the relatedness of a topic towards the pre-defined topic and the existence or not of an arguable reason, where the evaluation of the argument toward its completeness take place. Considering the complexity of the proposed methodology and the comparison with state-of-the-art baselines \cite{Mohammad2017StanceTweets, Goudas2014ArgumentMedia, Roitman2016OnTopics}, the presented results that can be characterized as promising. The entire process is characterized as information retrieval task, thus the learning-to-rank approach was adopted and the metric precision at k was used, which indicates the precision among the k top results of the retrieval. Considering the three distinctive sub-tasks that consist the claim-oriented tweet retrieval task have an increasing complexity, as it is pointed out through the report of the IAA, where topic-relevance reached 90.1\%, clear stance 78.2\% and detection of arguable reason 75.2\%. 

\begin{table}[]
\centering
\resizebox{\textwidth}{!}{%
\begin{tabular}{|l|l|l|l|l|l|l|}
\hline
Author & Scope & Task & Algorithms & Metric & \multicolumn{2}{l|}{Score} \\ \hline
Bosc et al. \cite{Bosc2016TweetiesMedia.}  & macro-analysis & support / attack & EOP + RTE & F1 support & \multicolumn{2}{l|}{0.17} \\ \cline{5-7} 
 &  &  &  & F1 attack & \multicolumn{2}{l|}{0.0} \\ \cline{4-7} 
 &  &  & LSTM & F1 support & \multicolumn{2}{l|}{0.20} \\ \cline{5-7} 
 &  &  &  & F1 attack & \multicolumn{2}{l|}{0.16} \\ \hline
Cocarascu and Toni \cite{Cocarascu2018CombiningDatasets} & macro-analysis & \begin{tabular}[c]{@{}l@{}} \\ support / \\ attack / \\ neither \end{tabular} & LSTM &  & dataset \cite{Tan2017SpotTwitter.} & dataset \cite{Guo2013LinkingMedia} \\ \cline{5-7} 
 &  &  &  & P & 0.59 & 0.97 \\ \cline{5-7} 
 &  &  &  & R & 0.97 & 0.90 \\ \cline{5-7} 
 &  &  &  & F1 & 0.73 & 0.94 \\ \hline
Ma et al. \cite{Ma2018ClaimTwitter} & {\begin{tabular}[c]{@{}l@{}}micro \& macro\\ analysis\end{tabular}} & {\begin{tabular}[c]{@{}l@{}} \\ topic relatedness, \\ support/attack, \\ arguable reason\end{tabular}} & SVM light & MAP & 0.59 & 0.50\\ 
\cline{5-7} 
 &  &  &  & P@5 & 0.53 & 0.51 \\ \cline{5-7} 
 &  &  &  & P@10 & 0.48 & 0.44 \\ \hline
 
\end{tabular}%
}
\caption{The different sub-tasks that have been accomplished in the context of relations identification task. MAP = Mean Average Precision, P@5 = Precision@5, P@10 = Precision@10. Rounding took place in order the results to be displayed in the same scale.}
\label{table_relations_identification}
\end{table}

In Table \ref{table_relations_identification}, a summarization of research papers on the task of relations identification in text derived from social media takes place. The first column of the table presents the authors of the paper, the second one interprets the scope of the task(s) according to the proposed definition and the task are presented in the third column. The next three columns present the technical details (algorithms, metric, score) for the implementation of each proposed method. In \cite{Cocarascu2018CombiningDatasets} the results for the two datasets that their proposed methodology has been tested are presented. It should be stressed that the scores are not directly comparable, as different research papers carry out different tasks; instead we should focus on the coexistence of different approaches and the level of difficulty of each one. 

Each one of the presented research papers exploits data derived from twitter, as the focus of this literature review paper is AM in social media. Web-derived text seems to thrive as a source of AM pipelines, including the task of relation identification, but the source of data usually is a more structured source of text, such as debate forums \cite{Lawrence2017DebatingArgument, Morio2018AnnotatingMining, Galassi2018ArgumentativeLearning, Eidelman2019ArgumentERulemaking}. Although the information found in social media are characterized as noisy text and it is far from an ideal scenario for AM, the constant generation of content allows to the researchers to conduct research including the time axis in order to understand users' behaviour \cite{Lai2018StanceDebate, Lai2017ExtractingOpinion} and evaluate their impact beyond the network \cite{Maynard2017AAnalysis, Cortis2017SemEval-2017News}. Users in social media platforms usually express emotions or quick messages with very little argumentation, however the introduction of argumentative features can enhance other NLP tasks \cite{Addawood2017StanceCase, Cocarascu2018CombiningDatasets}. Both micro \cite{Schulz2018Multi-TaskSettings} and macro \cite{Lawrence2017UsingERulemaking} analysis have the attention of the research community, whereas they have been approaches that combine them \cite{Morio2018AnnotatingMining, Galassi2018ArgumentativeLearning}. Another research topic that has gained the interest of the research community is the reconstruction of implicit warrants, although the existing research papers \cite{Rajendran2016ContextualReviews, Rajendran2018IsDataset, Habernal2018TheWarrants} do not utilize social media as source.  

\subsection{Stance Detection}
\label{subsec:StanceDetection}

In contrast to relations’ identification task, stance detection is a popular task among researchers in the NLP community, either as autonomous and independent task either as a part of an extensive pipeline. Stance detection is thriving even in the challenging environment of social media and Twitter is often used as a source of information. 

Stance detection is related to many sub-fields of the wider NLP area, such as sentiment analysis, textual entailment, topic extraction and AM. In the context of AM in social media, we define stance detection as the task of automatically determining the attitude of the author towards a specific topic exploiting any kind of information that can be collected. The stance can be determined either exclusively by the content of the text either from combination of features that are capable to reveal specific characteristics such as argumentativeness \cite{Addawood2017StanceCase} or network communities \cite{Lai2018StanceDebate, Grcar2017StanceReferendum}.

The main difference between opinion mining and stance detection as it is expressed in AM pipelines lies in the concept of data aggregation from the wider environment towards the final outcome. Stance detection is considered as the final part of the pipeline that exploits the findings of the previous steps, rather than a stand-alone task.

The main difference between opinion mining and stance detection as it is expressed in AM pipelines lies in the concept of data aggregation from the wider environment towards the final outcome. The term wider environment applies on both combination of sources and tasks \cite{Lytos2018ArgumentationKnowledge}, as web-generated data and especially social media offer an excellent environment for sentiment analysis, but a poor one for argumentation or opinion mining, thus stance detection is considered as the final part of the pipeline that exploits the findings of the previous steps, rather than a stand-alone task.

As the research community has shown great interest in the task of stance detection, it is impossible to present each research paper, rather we decided to focus on the research methodologies that either aggregate data from different sources either have the ability to be used as part of a bigger system.  
Similar to relations' identification task, we also provide a mathematical expression of the stance detection, influenced by \cite{Liu2012SentimentMining} on the definition of opinion mining. Stance detection is expressed as the quintuple:

\[ (h_i, s_{\text{ijkl}}, d_j, r_k, t_l) \]

where \textit{h\textsubscript{i}} is the person who holds a specific stance \textit{s\textsubscript{ijkl}} for a specific debate \textit{d\textsubscript{l}}, justified by a rationale \textit{r\textsubscript{k}} in a specific time \textit{t\textsubscript{l}}. The rationale of the stance for a specific debate can be assessed for its quality \cite{Dusmanu2017ArgumentSources,Addawood2016WhatMedia} through a variety of sub-tasks, such as facts identification, evidence recognition, source classification and reasoning evaluation. 

\begin{table}[]
\centering
\resizebox{\textwidth}{!}{%
\begin{tabular}{|l|l|l|l|l|}
\hline
Author & \# classes & Algorithm & Metric & Score \\ \hline

Zarella and Marsh \cite{Zarrella2016MITREDetection} & Favor/against/neither & RNN & F1 & 0.68 \\ \hline
Mohammad et al. \cite{Mohammad2017StanceTweets} & Favor/against/neither & linear-kernel SVM & F-micro & 0.70 \\ \cline{4-5} 
 &  &  & F-macro & 0.59 \\ \hline
Wei et al. \cite{Wei2016PkudblabDetection} & \begin{tabular}[c]{@{}l@{}}Favor/against/neither\end{tabular} & Neural network & F1 & 0.56 \\ \hline
Wei et al. \cite{Wei2018Multi-TargetNetwork} & \begin{tabular}[c]{@{}l@{}}Favor/against\end{tabular} & Neural network & F1 & 0.71 \\ \hline
Ebrahimi et al. \cite{Ebrahimi2016WeaklyBootstrapping} & Favor/against/neither & Linear-kernel SVM & F macro & 0.57 \\ \hline
Johnson and Goldwasser \cite{Johnson2016IdentifyingTwitter} & Favor/against & Probabilistic Soft Logic & A & 0.86 \\ \hline
Lai et al. \cite{Lai2018StanceDebate} & Favor/against & SVM & F-macro & 0.90 \\ \hline
Addawood et al. \cite{Addawood2017StanceCase} & Favor/against/neutral & SVM & P & 0.90 \\ \cline{4-5} 
 &  &  & R & 0.90 \\ \cline{4-5} 
 &  &  & F1 & 0.90 \\ \hline
\end{tabular}%
}
\caption{The results of the supervised and weakly-supervised ML approaches that have been followed for the stance detection in social media text. Fully supervision on \cite{Zarrella2016MITREDetection}, \cite{Mohammad2017StanceTweets}, \cite{Addawood2017StanceCase}. Weakly supervision on \cite{Wei2018Multi-TargetNetwork}, \cite{Ebrahimi2016WeaklyBootstrapping}, \cite{Johnson2016IdentifyingTwitter}, \cite{Lai2018StanceDebate}. The \cite{Zarrella2016MITREDetection}, \cite{Mohammad2017StanceTweets}, \cite{Wei2016PkudblabDetection}, \cite{Ebrahimi2016WeaklyBootstrapping} and \cite{Addawood2017StanceCase} are applied on the same dataset. RNN = Recurrent Neural Network   }
\label{TableStanceDetection}
\end{table}

The sixth task of SemEval-2016 \cite{Mohammad2016SemEval-2016Tweets} introduced the shared task of stance detection in tweets, providing a significant boost in the field as new methodologies were suggested and the constructed dataset was also used in later research. The shared task consisted of two parts regarding the supervision framework to be followed (fully-supervised, weakly-supervised). As the constructed dataset was used afterwards from the completion of the task, more modern approaches have surpassed the top performances described in the task.

The best-performing system of the competition \cite{Zarrella2016MITREDetection} proposed a recurrent neural network capable to extract information from unlabeled datasets using word embeddings. The use of word embeddings as features was also critical in \cite{Mohammad2017StanceTweets}, where a simpler linear SVM algorithm achieved F-score up to 0.70, surpassing the previously highest score of 0.68. Apart from the use of word embeddings, the presence or absence of the target of interest in the tweet improved the results of the algorithm. One more improvement in the same dataset for the same task was achieved by Wei et al. \cite{Wei2018Multi-TargetNetwork} reaching the F1 to 0.71, where an end-to-end neural model was proposed which makes better use of target information.  
  
Considering the results for the weakly-supervised framework as described in \cite{Mohammad2016SemEval-2016Tweets}, those are significantly lower as no training data are provided for the topic that is researched, thus the developed methodologies rely heavily on techniques that can transfer knowledge from different topics. The submission with the highest performance for the task achieved 0.56 F1 \cite{Wei2016PkudblabDetection} and proposed a convolutional neural network including a modified softmax layer able to perform three class classification, although the training is consisted of two classes. A weakly-supervised approach exploiting the network structure information proposed from Ebrahimi et al. \cite{Ebrahimi2016WeaklyBootstrapping} improved the previously best score and reached 0.57 F1.  

Weakly supervised approaches exploiting social media features for political stance detection were also adopted in Lai et al. \cite{Lai2018StanceDebate} and in Johnson and Goldwasser \cite{Johnson2016IdentifyingTwitter}. The former is focused on the Italian referendum in 2016 and employs a holistic approach for the stance detection adopting a diachronic perspective for the user's stance including twitter-specific features and social network communities, achieving 0.90 f-micro with the use of svm. A similar approach was also followed in \cite{Johnson2016IdentifyingTwitter}, able to capture both the content and the social context through linguistic patterns reaching 0.86 accuracy. The novelty in their approach lies on the absence of manual annotation, as the annotation of the political stances took place with the use of ISideWith.com.

In the work of Addawood et al. \cite{Addawood2017StanceCase}, an advancement of a previously established scheme \cite{Addawood2016WhatMedia}  is proposed capable to carry the task of stance detection resulting in a F1 score of 0.93 with the use of the Decision Tree algorithm. In this paper, the introduction of argumentativeness as a feature take place increasing significantly the performance of the algorithm. The findings of the paper indicate that argumentativeness features are the most informative ones for the successful categorization of favor and neutral categories, stressing the importance for introducing AM techniques in different text mining tasks.

Table \ref{TableStanceDetection} presents the algorithms, the metrics and the score selected research papers have achieved. In the second column the possible classes that are provided to the classification algorithms are depicted, there is either a binary approach (favor, against) either the ‘neither’ option is included. Considering the algorithms that are used, SVM and neural networks algorithms are used, apart from \cite{Johnson2016IdentifyingTwitter} where a probabilistic soft logic was adopted. In the last two columns, the score that has achieved measured with different metrics are presented. As it is expected fully supervision ML approaches achieve higher results when they are compared to weakly supervision ML approaches, when both applied on the same dataset. Apart from the ML approach, defining the number of possible classes plays a role in the performance of the algorithms, as the binary approach achieves normally higher results.

\subsection{Reliability-related tasks}
\label{subsec:Reliability-related}

The wide use of Twitter in combination with its public nature have established it as the most appropriate social network for studying a variety of tasks related to virality, such as viral marketing, rumour diffusion, event and fake news detection. The majority of the proposed methods are relied to social network analysis, exploiting the metadata offered by the social network (friends, followers, time of publishing, etc.). As the scope of this paper is neither an in-depth review of rumour detection techniques and methods, nor the evidence identification for claims in any kind of text, we are going to focus on research work that connects argumentativeness with the evaluation of the argument's reliability in text derived from social media.

The Twitter platform is often used as a mean of expressing arguments for controversial topics, some of them are efficiently supported with facts and evidence from reliable sources, whereas in some other cases instead of backing their claims, they simply express feelings or unsupported allegations. The constant data generation and rapid pace of news flow create a chaotic environment with limited time for claims to evaluated and facts to be assessed. Due to the environment that has been created, where users express opinions and views in real time without using sophisticated or pretentious vocabulary, a unique opportunity is raised for argument evaluation on various political and social issues. The automatic evaluation of arguments has the potential to reduce the incidents of rumour spreading, the faster detection of fake news and eventually the improvement of the quality of public political discourse.

An essential part of a complete argument is the sufficient backing of the original claim, either in the form of premises either in the form of backing with evidence and facts presentation. A simple, but robust structure of claim and supporting evidence is adopted in \cite{Addawood2016WhatMedia} for the classification of arguments' evidence, where the ultimate step of the proposed pipeline is the classification of evidence to six different categories. The proposed pipeline of Dusmanu et al. \cite{Dusmanu2017ArgumentSources} contains two tasks that are related to the reliability of the expressed argument, the distinction of factual information from opinions and the source identification from a pre-defined list. For both the tasks the use of all the available features boost the results of the classification.    

In the work of Konstantinovskiy et al. \cite{Konstantinovskiy2018TowardsDetection}, where the objective is the construction of a reliable fact-checking mechanism, both an annotation scheme and an automated claim detection method is proposed. Two different approaches are presented, the binary model of claim/no claim and a multi-class classification with seven categories describing the claim. The proposed methodology (the binary model) overcomes previously established mechanisms (Claimbuster, ClaimRank) in terms of F1, as it achieves F1 0.83, while the multi-class classification displays the impressive 0.70 F1-micro, and in terms of macro average it achieves 0.48 F1-macro.

\begin{table}[]
\resizebox{\textwidth}{!}{%
\begin{tabular}{|l|l|l|l|l|l|}
\hline
\textbf{Author} & \textbf{Task} & \textbf{\# classes} & \textbf{Algorithm} & \textbf{Metric} & \textbf{Score} \\ \hline
Addawood et al. \cite{Addawood2016WhatMedia} & Evidence Classification & 6 & SVM & F1-macro & 0.83 \\ \hline
Dusmanu et al. \cite{Dusmanu2017ArgumentSources} & Factual vs opinion & 2 & LR & P & 0.81 \\ \cline{5-6} 
 &  &  &  & R & 0.79 \\ \cline{5-6} 
 &  &  &  & F1 & 0.80 \\ \cline{2-6} 
 & Source Identification & NA & str match + h. & P & 0.69 \\ \cline{5-6} 
 &  &  &  & R & 0.64 \\ \cline{5-6} 
 &  &  &  & F1 & 0.67 \\ \hline
Konstantinovskiy et al. \cite{Konstantinovskiy2018TowardsDetection} & Claim detection & 2 & LR & P & 0.88 \\ \cline{5-6} 
 &  &  &  & R & 0.80 \\ \cline{5-6} 
 &  &  &  & F1 & 0.83 \\ \cline{3-6} 
 &  & 7 & LR & P-micro & 0.71 \\ \cline{5-6} 
 &  &  &  & R-micro & 0.73 \\ \cline{5-6} 
 &  &  &  & F1-micro & 0.70 \\ \cline{5-6} 
 &  &  &  & P-macro & 0.61 \\ \cline{5-6} 
 &  &  &  & R-macro & 0.44 \\ \cline{5-6} 
 &  &  &  & F1-macro & 0.48 \\ \hline
\end{tabular}%
}
\caption{The results of the supervised ML approaches that have been followed for reliability-related tasks in AM, in text derived from social media. For the task of source identification a rule-based approach is followed. str match = string matching. h = heuristic algorithm}
\label{TableReliability}
\end{table}

In Table \ref{TableReliability} a summarization of the research work that accomplishes reliability-related tasks in the context of AM in text derived from social media is presented. Four tasks have been recognized in this category and the results of the different approaches are heavily relied on the number of the alternative classes. The 0.83 F1-macro that is achieved in \cite{Addawood2016WhatMedia} is impressive if we consider that there are six available classes, and for a similar task with seven available classes the F1-macro is 0.48 in \cite{Konstantinovskiy2018TowardsDetection}. The difference could be merely explained through the exploitation of more features in \cite{Addawood2016WhatMedia} in comparison to \cite{Konstantinovskiy2018TowardsDetection}, where not that advanced features were used. 

Besides the aforementioned research, there have been some approaches on connected reliability and evidence with argument's strength, but they are not applied in social media text. For example, a research work that uses argumentativeness as a feature, apart from \cite{Addawood2017StanceCase}, takes place in Cocarascu and Toni \cite{Cocarascu2018CombiningDatasets} for the task of deceptive reviews detection, leading to an improvement of the prediction algorithms. It has to be noticed that the exclusive use of argumentative features without topic modelling or the use of additional features cannot surpass the baseline. The work of Park and Cardie \cite{Park2014IdentifyingComments, Park2015ConditionalComments} is another great example of combining argumentation with evidence classification, suggesting three different categories of justifications in online user comments, but as in \cite{Cocarascu2018CombiningDatasets}, both papers exploit more structured forms of arguments. Similarly, the task of context dependent claim detection \cite{Rinott2015ShowDetection, Levy2014ContextDetection} utilizes hundreds of Wikipedia articles, but it has not been applied in social media text.  

Another task that is related with both the quality evaluation of the argument and the in-depth analysis in macro-scale is the enthymeme reconstruction. Although the task has not been applied yet in data derived from social media, at least in our knowledge, it has been tested in web-generated data \cite{Rajendran2016ContextualReviews, Rajendran2018IsDataset}  and we strongly believe that text from social media offer an excellent testing ground for the identification of implicit supporting claims. The task of reasoning comprehension as it has been presented in \cite{Habernal2018TheWarrants} has the potential to be applied in social media and advance the wider field of AM, but the complexity of the proposed methods raises concerns for the transferability of the task to new environment and un-trained annotators.

\section{Existing Tools in AM} \label{sec:Tools}

The increasing interest in AM has increased the need for suitable tools, such as grammar parsers, sentiment lexicons, software for boosting the manual annotation tasks, and tools capable to automatically extract arguments from natural language. In the area of NLP, there is a wide range of available tools covering various aspects and addressing different challenges. However, there is lack of standardization, accessibility, and acceptability of the existing tools, due to proprietary formats developed for modeling of natural language in different domains.

In subsection \ref{subsec:ToolsGeneralPurpose}, we provide a synopsis of some of the most prominent tools in the area of NLP and how they could improve the process of accomplishing AM tasks. In the second subsection \ref{subsec:Tools_AMspecific}, we narrow down the area of interest and focus on tools that are specified for the task of AM. 

\subsection{General-purpose NLP Tools} \label{subsec:ToolsGeneralPurpose}

The annotation process is of major importance in any NLP system, thus different tools have been proposed following different approaches. The first introduced web-based opensource annotation tool is BRAT \footnote{http://brat.nlplab.org/index.html} \cite{Stenetorp2012BRAT:Annotation}, which is based on the STAV text annotation visualizer \cite{Stenetorp2011BioNLPResources} and is characterized by the wide variety of tasks that can be accomplished and scientific work that has been conducted using it. It has been adopted in different fields like visualization, entity mention detection, event extraction, coreference resolution, normalization, chunking, dependency syntax, and meta-knowledge annotation.

A tool following the approach of BRAT is WebAnno \cite{EckartDeCastilho2016AStructures} which has kept the web interface and visualization capabilities of BRAT and modified the server layer. WebAnno has improved specific weaknesses of BRAT focusing on user and quality management with the addition of monitoring tools and interfaces for crowdsourcing. Currently, WebAnno is in version 3.0 and also offers a web-instance\footnote{https://webanno.sfs.uni-tuebingen.de/} through the CLARIN-D infrastructure. Both BRAT and WebAnno are an open and live project that  could easily be modified in order to include tasks in the sphere of AM, such as argument detection, relation identification or reasoning evaluation.

The construction of graphs for text annotation is followed in GraPAT \cite{Sonntag2014GraPAT:Annotations} covering different tasks like sentiment analysis, argumentation structure, rhetorical text structure, and natural visualization of the annotation process. The initial goal for the development of GraPAT was to increase IAA and maximize the automation of annotation without neglecting neither the variability nor the annotation speed/comfort. GraPAT can be considered as the successor of RSTTool \cite{ODonnell2000RSTToolTheory}, as it maintains the principles of RST annotation, enriching it with more capabilities like sentiment analysis and argument structure annotation model.  

GATE \cite{Cunningham2013GettingAnalytics} has dominant presence in the wider field of text engineering offering numerous tools and capabilities from simple tasks (e.g. information extraction, named-entity, etc.), to modifications for cutting-edge technologies, such as cloud-enabled software and social media analysis. Regarding the argument annotation task, GATE offers Teamware \cite{Bontcheva2013GATEFramework}, a web-based software suite which provides the environment for collaborative annotation and curation. GATE Teamware stands out as the only annotation tool, to the best of our knowledge, which supports execution of an automatic NLP system, before manual annotation. 

The DiGAT tool \cite{Kirschner2015LinkingPublications} has been developed alongside with an annotation scheme and a graph-based inter-annotator agreement measure based on semantic similarity. Similarly to GraPAT, DiGAT also relies on graph structures for the annotation process, aiming at simple and accurate annotation of relations between entities in long text.  

The establishment of the TextCoop platform alongside the Dislog language is presented in \cite{Saint-Dizier2012ProcessingPlatform}. TextCoop is the only tool in this subsection that follows a logic-based approach, heavily influenced by RST, modeling the conclusion as a nucleus and the support as a satellite. As it is described, TextCoop offers a functional web interface, however, to the best of our knowledge, this is not provided, yet. Another tool that is in a similar status with TextCoop is TURKSENT \cite{Eryiit2013TURKSENT:Media}, a manual annotation tool with multi-lingually capabilities aiming at automatic sentiment analysis of text derived from social media. Its mentioned web-based interface does not seem to be available, yet.

\begin{table}[]
\resizebox{\textwidth}{!}{%
\begin{tabular}{|l|c|c|c|c|}
\hline
Tool & Web UI & Manual Annotation & Arg Retrieval & Arg Evaluation \\ \hline
WebAnno \cite{Yimam2013WebAnno:Annotations} & Yes & Yes &  &  \\ \hline
BRAT \cite{Stenetorp2012BRAT:Annotation} & Yes & Yes &  &  \\ \hline
GraPAT \cite{Sonntag2014GraPAT:Annotations} & Yes & Yes &  &  \\ \hline
DiGAT \cite{Kirschner2015LinkingPublications} & Yes & Yes &  &  \\ \hline
MARGOT \cite{Lippi2016MARGOT:Mining}  & Yes &  & Yes & Yes \\ \hline
OVA+ \cite{JanierMathilde2014OVA+:Interface} & Yes & Yes &  &  \\ \hline
TOAST \cite{SnaithMark2012TOAST:Implementation} & Yes &  & Yes & Yes \\ \hline
GATE Temware \cite{Bontcheva2013GATEFramework} & Yes & Yes &  &  \\ \hline
Args \cite{Wachsmuth2017BuildingWeb} & Yes &  & Yes & Yes \\ \hline
ArgumenText \cite{Stab2018ArgumenText:Sources} & Yes &  & Yes & Yes \\ \hline
Rationale \cite{vanGelder2007TheRationaleTM}  & Yes & Yes &  &  \\ \hline
\end{tabular}%
}
\caption{A summarization of the existing NLP tools that can enhance the process of AM. The table includes tools in the wider NLP area which can be integrated at any stage of an AM pipeline.}
\label{tableTools}
\end{table}

The Stanford CoreNLP toolkit \cite{Manning2014TheToolkit} receives great acceptability from the NLP community as it offers a broad range of grammatical analysis tools, different APIs for the majority of the programming languages, and the ability to run as a simple web service. However, a specific tool for AM-related tasks is not yet developed. Among the existing tools offered by the toolkit, the Stanford OpenIE is more closely related with AM, as it enables the extraction of relation tuples out of binary relations. Two more tools included in the toolkit and can be used in an AM pipeline is the Stanford Relation Extractor, which finds relations between two entities located by the Stanford Named Entity Recognizer, and the Neural Network Dependency Parser, a dependency parser that establishes relationships between "head" words and "modifier" words in a sentence. The web interface provided by the toolkit\footnote{http://corenlp.stanford.edu/} offers up to ten different annotators and the visualization of the schemes has been realized using the BRAT software.

The majority of the aforementioned tools are on-going, open-source projects that can be modified in order to carry out AM-related tasks, whereas some others (GraPAT, DiGAT) are graph-based annotation tools that are used for identifying the relations between chunks of text. Other functionalities such as sentiment evaluation and name entity recognition can boost AM related tasks, as sentiment features are used in the majority of the existing research papers in the area as Table \ref{FeaturesTable} illustrates. If any of the subtasks that are included in the AM pipeline can be executed automated and reliable through a software tool, then this tool should be exploited.

\subsection{Argument search, retrieval and automatic annotation} \label{subsec:Tools_AMspecific}

In this sub-section, we present the tools that have been designed to enhance the argumentation process. Some of those tools have been implemented to boost the annotation step, others offer an argumentation search engine and there are tools capable to automatically grade an argument or even perform the entire process of AM.   

The Centre for Argument Technology\footnote{http://www.arg-tech.org/} has produced a series of tools covering different aspects of AM. The latest developed tool is OVA\footnote{http://ova.arg-tech.org/} \cite{JanierMathilde2014OVA+:Interface} and has replaced to a certain extent the tool of Araucaria \cite{REED2004Araucaria:Representation}. Arivina\footnote{http://arvina.arg-tech.org/} \cite{Snaith2010MixedDeliberation}, the successor of MAgtALO \cite{Reed2007DialogicalDebates}, offers a dialogue system implementing the concept of mixed initiative argumentation, where human players and agents debate having equal levels of participation. The process for the calculation of the acceptability semantics on structured argumentation frameworks is completed through TOAST\footnote{http://toast.arg-tech.org/} \cite{SnaithMark2012TOAST:Implementation}. 

Probably the most influential achievement of the considered organization is the establishment of the Argument Web\footnote{http://www.argumentinterchange.org/} \cite{Reed2017TheArgumentation}, a repository in cooperation with a series of tools, systems and services, such as AIFdb, the main search interface for the Argument Web. ArguBlogging\footnote{http://www.argublogging.com/} \cite{Bex2014ArguBlogging:Web} materializes the concept of crowdsourcing in argumentation by capturing arguments that take place in online platforms (tumblr and Blogger are supported) and provides them as feed to the Argumentat Web. Concerning the educational aspect of argumentation, Argugrader\footnote{http://www.argugrader.com/} offers automatic grading and provides detailed feedback to students regarding successful or not construction of arguments.

A prototype argument search framework is proposed in \cite{Wachsmuth2017BuildingWeb} able to carry out the entire process of an argumentation search engine, from user query and argument retrieval to ranking and presentation of arguments\footnote{http://www.arguana.com/args}. The argument search engine is relied on pre-structured  arguments from a defined list of debate portals, and a standard mapping process takes place in order to convert the  concepts that characterize each argument in the different debate portals to the common argument model.

A system able to utilize the heterogeneity and big volume data is implemented in Stab et al. \cite{Stab2018ArgumenText:Sources}, under the name ArgumenText. The ArgumenText uses
400 million plain-text documents from different sources and deploys a series of technologies in order to construct a solid pipeline able to materialize a sequence of sub-tasks that eventually present ranked pro and con arguments through a web interface.      

At this moment we are aware of only one tool that accomplishes the complete task of automatic annotation in terms of AM. MARGOT \cite{Lippi2016MARGOT:Mining} is built on the foundations of Lippi and Torroni \cite{Lippi2015Context-independentMining} and extends the previous established model by including the task of evidence detection and providing a web interface\footnote{http://margot.disi.unibo.it/}. The syntactic structures that are followed in argumentative discourse is the fundamental idea on which the tool was built. The model implemented for AM involves two binary classification problems, the argumentative sentence detection and the argument components boundaries detection. The former is addressed with a combination of tree kernel and bag-of-words, whereas for the latter SVM-HMM with bag-of-words, part-of-speech, lemma and named entity are employed. In overall, the tool achieves acceptable evaluation scores, regarding the complexity of the task, but there is still room for improvement in covering various domains. 

Another tool that is very similar to MARGOT, sharing the same interface\footnote{http://155.185.228.137/claudette/}, is CLAUDETTE \cite{Lippi2018CLAUDETTE:Service}, an on-line platform addressing possible unfair or abusive terms of service. The platform realizes a 2-step algorithm including the binary task of detecting an unfair clause and if the first step is positive, then a classification task through 8 possible categories takes place. For this classification, a combination of eight SVMs exploiting lexical features is used with fair results in all used metrics.

On industry level, Austhink Pty Ltd. is continuously developing software tools aiming at the improvement of the general reasoning and argumentation process through training sessions. The two most successful software tools are Rationale \cite{vanGelder2007TheRationaleTM,Sbarski2008VisualizingStructure} and bCisive \cite{Marriott2011Hi-TreesLayout}, with both of them offering online limited free versions\footnote{https://www.rationaleonline.com/editor/}\footnote{https://www.bcisiveonline.com/editor/}. The former supports a series of activities in the area of AM, such as construction, visualization and mapping of arguments, while the latter focuses on providing support to business decision through hypothesis and decision maps. Rationale is considered as the successor of Reason!Able \cite{Gelder2000LearningApproach}.

A synopsis of the existing NLP tools that have been discussed in this section is presented in Table \ref{tableTools}. The existing tools can be classified into three categories, tools that aid the manual annotation process, general-purpose NLP tools and tools that offer an entire mechanism for argument search, retrieval and evaluation. It has to be underlined the evaluation of the argument differs to the different approaches, as in \cite{Lippi2016MARGOT:Mining} the number of claims and premises are presented, in \cite{SnaithMark2012TOAST:Implementation} the weight of the argument is calculated and in \cite{Wachsmuth2017BuildingWeb, Stab2018ArgumenText:Sources} the arguments are categorized as pro and con.  

\section{Proposed conceptual framework and future directions}
\label{sec:FrameworkFuture}

As every scientific field that is in a premature level, AM has to take steps towards its establishment in the wider research community of NLP by invoking the interest to a greater audience through more applications of AM in a wider spectrum of scenarios. Towards this vision, there are specific challenges and concerns that have to be faced and technologies that have to be tested. In the remainder of this section, we propose a conceptual framework able to capture the needs of AM in social media and boost NLP-related tasks, and present some promising approaches that can significantly contribute in the AM process. Our proposed conceptual framework is in our knowledge the first pipeline capable to connect various NLP tasks between them having in prominent position the AM tasks in noisy environment. Previously proposed pipelines are either focused on the task-at-hand \cite{Addawood2016WhatMedia, Al-Khatib2016Cross-DomainSupervision, Park2014IdentifyingComments} either do not present any connection with other NLP-related tasks \cite{Lippi2016ArgumentationMining}.

\subsection{Proposed conceptual framework for AM in social media}
\label{subsec:ProposedConceptualArchitecture}

The two most dominant characteristics of text derived from social media is the short length and the lack of defined norms. Considering the fact that the typical length of a tweet is less than 50 characters, the definition of argument boundaries in many cases is not feasible, as an argumentative tweet can hardly contain information unrelated to the major claim. On the other hand, the already unstructured nature of text data in combination with the massive use of jargon and emoticons establish an environment where any lexical rule is really challenging to be applied, as either has to be specific for each case study either loosely defined in order to include different cases. Both assumptions would probably lead to lack of transferable knowledge, a crucial objective for almost every proposed methodology.

A possible criticism for the lack of the boundaries-definition task as defined in \cite{Lippi2016ArgumentationMining} is the increase of the upper limit in Twitter to 280 characters from 140, as well as the existence of various social media including forums such createdebate.com where the norm indicates well-structured arguments and length significant longer than 50 characters. However, the great dominance of Twitter in socio-political issues, which is also boosted from the online presence of political leaders, combined with the shrinking of general-purpose forums have formed an environment which the use of social media data seems to be the only source of web-generated data in the near future.

\begin{figure}[t]
\includegraphics[width=12cm, height=9cm]{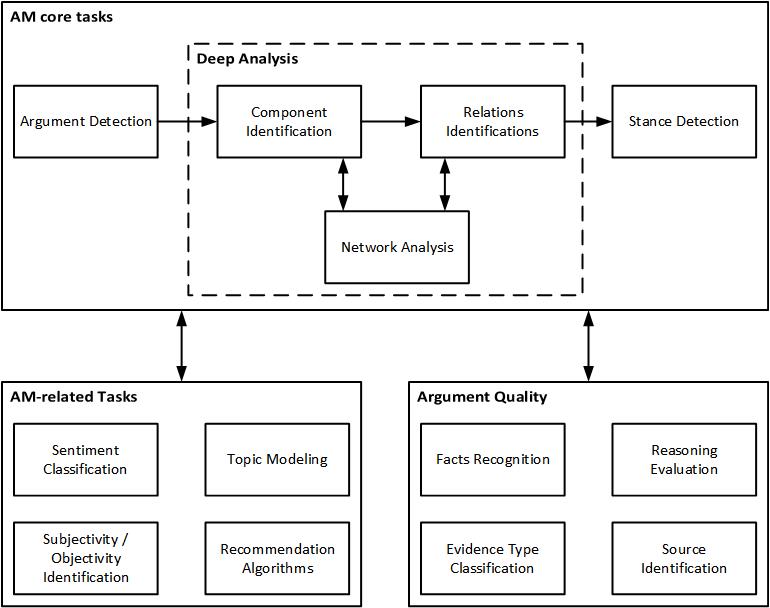}
\centering
\caption{The Proposed Conceptual Architecture for AM} 
\label{fig:ArgMinScheme}
\end{figure}

The increase use of social media in the wider area of AM as a source of information highlights the need for the definition of a new scheme devoted for the specialized procedures in the field of AM in social media. The proposed conceptual architecture, as it is depicted in Figure \ref{fig:ArgMinScheme}, contains three main components and each one includes distinct tasks that could form a second focused pipeline, but could also exist independently from the other tasks. The first component contains the core tasks of an AM procedure, the tasks that make the heart of the pipeline and can be applied in different text data, from official political speeches to tweets and comments in products or services reviews. The other two components include tasks that are involved somehow in the process of evaluation the reliability of the text. The additional components should not be underestimated and should be considered of equal importances compared to the AM core task, as the interest of the scientific community is constantly increasing for tasks such as fact checking, evidence and fake news detection, even though the connection between AM and reliability-related tasks is not yet very solid.

The first step of the first component in the proposed conceptual architecture is the task of argument detection, where the identification of a sentence as argumentative or not takes place. A significant amount of work intentionally ignores the step of argument detection \cite{Stab2014AnnotatingEssays, Lai2018StanceDebate}, as argumentative rhetoric is a prerequisite for the evaluation of persuasive essays or political debates. When collecting heterogeneous data from social media the detection of argumentative text is unavoidable, as not all users intend to persuade for or against a discussed topic, but simply express a reflection, a feeling or a question. The task of argument detection is considered as an essential step for the AM pipeline in social media, as the following steps of pipeline is not possible to be completed if is missing.

A great amount of research work is focused on the identification of the components that construct the argument, especially the early attempts of argumentation (see section \ref{subesec:LogicalSchemes}), presenting as the ultimate goal the successful analysis of the argument, emphasizing in the reasoning concept behind the structure of the argument. The concept of in-depth analysis of an argument through the identification of the argument's components and the discovery of the underlying relations between them has been adopted and developed in the field of AM in social media, adjusting to the new environment including the relations and the interaction between the entities of the network. The two main tasks that fall under this category is the relation-based AM \cite{Carstens2017UsingProblems} and the enthymeme reconstruction \cite{Rajendran2016ContextualReviews}. Both tasks provide a useful insight for the structure of the argument, but instead of trying to evaluate its impact, they focus on the efficient use of small datasets \cite{Rajendran2018IsDataset, Cocarascu2018CombiningDatasets} or they are designed to aid the task of stance detection \cite{Rajendran2016ContextualReviews}.

Earlier in the text we characterized Opinion Mining as the predecessor of AM, mostly because the terms stance detection and opinion mining can be used interchangeably. Stance detection is the final step of the proposed framework and it is a task that is heavily depended from the previous steps of the pipeline, as the nature of social media data demands a significant pre-process procedure. This procedure in AM pipeline is expressed through the detection of argumentative tokens and the identification of relationship between the components of the argument (explicit or implicit).   

The constant generation of data in social media has raised significant concerns for the quality of information that is shared and read in social media. The connection between argument in social media and reliability has not yet discovered in depth, comparing with the amount of work that is dedicated in the areas of rumour and fake news detection. However, tasks such as evidence type classification, source identification and facts recognition have emerged in the area of AM, raising the awareness of the connection between the expressed argument and its reliability. 

In the proposed conceptual architecture for AM in social media we devote a unique component of the pipeline in reliability-related tasks, in order to stress its importance and the room of the development that exists in the area. AM reliability tasks are not considered as core tasks of the AM pipeline, but they can enhance the procedure especially when applied in arguments derived from social media, as the backing of the claims is many times inaccurate or it is based on rumours or hoaxes. Other tasks that can be assisted by the progress in the field of AM and integrate parts of the proposed AM core tasks is sentiment classification \cite{Mohammad2017StanceTweets}, subjectivity/objectivity identification, topic modeling \cite{Wachsmuth2017AAnalysis}, and recommendation algorithms \cite{Karimi2018NewsAhead}.

Different tasks can be easily executed and combined through the components described in our proposed architecture, as it is both detailed and easily modifiable. The example provided in 3.1 concerning the Apple/FBI encryption debate (\textit{RT @ItIsAMovement "Without strong encryption, you will be spied on systematically by lots of people" - Whitfield Diffie}), can be assessed for its argumentativeness nature, the relations that expressed through the retweet and the mention, its stance towards the discussed debate, its completeness and integrity, while other NLP-related tasks can also be enhanced from the findings of the above tasks. Our intentions on our proposed framework is to be regarded and used as a mean for enhancing various NLP tasks with the use of argumentativeness features.

\subsection{Future directions: semi-supervision and background knowledge}

Handling data with great volume and variety is not an easy task, especially if we consider the well-known \textit{"gold rule"}, where manual annotation is required for a subset of data. The process of manual annotation is labor intensive and time consuming, whereas possible use of unsupervised machine learning algorithms could solve the problem of lack of trained annotators. The need for novel algorithms and techniques is emphasized also in previous review papers \cite{Mochales2011ArgumentationMining, Lippi2016ArgumentationMining} and although there is an evident trend towards unsupervised \cite{Boltuzic2015IdentifyingSimilarity, Duan2019AnDetection}  or semi-supervised ML algorithms \cite{Park2015ConditionalComments,Sardianos2015ArgumentNews, Al-Khatib2016Cross-DomainSupervision, Shnarch2018WillMining} with notable performance, they could be further improved, as there has not been extensive work neither on the use of suitable features nor on the design of argument schemes. A recent literature review from Silva et al. \cite{Silva2016ALearning} presents the trends in semi-supervised learning for tweet sentiment analysis and it can be used as a point of reference also in the field of AM.

Deep learning techniques are able to handle a great volume of data in an unsupervised or semi-supervised way and they have achieved break-trough results in NLP field. Deep learning has been applied \cite{Schulz2018Multi-TaskSettings,Habernal2017ArgumentationDiscourse,Rajendran2018IsDataset} in AM, but does not seem to overpass other ML algorithms, mainly because of the limited available datasets, however more research should take place in order safe conclusion to be drawn. 

In the fields of Sentiment Analysis and Opinion Mining there is a major trend towards deep learning, which is adopted in multidisciplinary fields and in various kinds of test cases \cite{Nguyen2018MultilingualEmbedding, Day2017DeepReview}. Concerning related applications in the AM field, several deep learning architectures are combined with unsupervised learning in the pretraining stage for feature extraction, in order to obtain important semantic features through this process. It is probably infeasible to completely solve such a complex problem, like AM, without the use of manual annotation, but these techniques could significantly reduce load from the labor intensive process of manual annotation.

The concept behind AM is finding the underlying reasoning of an expressed opinion, not only identifying if this opinion is positive or negative. In human reasoning, this process is accomplished naturally by combining a-priori knowledge towards a specific subject and knowledge on the ethos and influences of the person who express an opinion. A similar approach is followed for irony detection \cite{Wallace2015SparseSentiment}, where previous work determined that contextualizing information is required \cite{Wallace2014HumansToo}. Oddly enough, an approach of that sort is yet to be followed in the field of AM, although combining multiple sources of information seems to be the natural flow of human reasoning for argument detection and classification. Exploiting background knowledge can be achieved through the use of semantic encoders mainly in models employing reinforcement learning.           

The studies in the field of opinion mining and sentiment analysis are focused on real-world scenarios combining different scientific domains like politics or social networking and their roots are originated from reviews, recommendation systems, and digital marketing. Most of the research on opinion mining aims at improving specific commercial applications, whereas AM architectures have not been tested thoroughly in such scenarios. In order AM to go one step further, it is crucial that more real-world scenarios will be employed combining background knowledge and information from multiple sources.

\section{Conclusion}
\label{sec:conclusion}

Argumentation Mining is an attempt for a deeper understanding of natural language, is the natural evolution of opinion mining, but instead of trying to understand what others think, the focus is on understanding why. The analysis of the human reasoning process is the ultimate goal of AM and it is acquired by exploiting the inherent structure of an argument through the identification of its distinctive parts implying the inferential process that is followed. 

The understanding of the human reasoning can offer unprecedented capabilities and achieve breakthrough changes in a wide spectrum of applications as information for the decision-making process can be retrieved. The existing work in the field indicates its great potential but we must consider the fact that AM is still in a premature stage and there are steps that are required for realizing human-level reasoning or at least to be able to interpret it at a sufficient level. The research community has focused on the modeling of the argument and in the suitable selection of features, whereas the selection of the ML algorithm seems not to play a crucial role for the accomplishment of the different tasks. Different approaches have been tested for modeling arguments from diagrams depictions to modifications of well-established theoretical models and no model excels in comparison with others, creating the need for establishing a flexible framework able to capture the needs of the different tasks that appear in AM problems.

In our review, we try to shed light to the Argumentation field, in order to provide a clear view of the wider area with a focus on automatic AM in social media text. We present different models that take place in previous research works and break them down to their core tasks. Inspired from the individual AM sub-tasks, we propose a complete conceptual architecture for AM that can be easily adopted and modified depending on the goals of every research work and we present the existing tools in both the wider NLP area and more specified tools for the task of AM.  

Argumentation Mining could stimulate a series of applications, where the evaluation and the classification of reasoning is essential, especially in web content, where both reliability and reasoning validity of a user holding a position are questionable. Tasks such as troll detection, knowledge retrieval, and information validation could be significantly benefited by the progress in AM. Cutting- edge techniques in human-computer interaction, opinion mining, and recommendation systems could adopt parts of an AM system and enhance their performance. The successful interpretation, evaluation, and taxonomy of arguments will eventually lead to human level reasoning machines, which can understand, evaluate, and eventually create knowledge.

\section*{References}

\bibliographystyle{model1-num-names}
\bibliography{references.bib}

\begin{thebibliography}{100}
\expandafter\ifx\csname url\endcsname\relax
  \def\url#1{\texttt{#1}}\fi
\expandafter\ifx\csname urlprefix\endcsname\relax\def\urlprefix{URL }\fi
\expandafter\ifx\csname href\endcsname\relax
  \def\href#1#2{#2} \def\path#1{#1}\fi

\bibitem{Toulmin2003TheArgument}
S.~E. Toulmin, {The Uses of Argument}, Cambridge University Press, Cambridge,
  2003.
\newblock \href {https://doi.org/10.1017/CBO9780511840005}
  {\path{doi:10.1017/CBO9780511840005}}.

\bibitem{Liebeck2016WhatFeld}
M.~Liebeck, K.~Esau, S.~Conrad, {What to Do with an Airport? Mining Arguments
  in the German Online Participation Project Tempelhofer Feld}, in: Proceedings
  of the Third Workshop on Argument Mining (ArgMining2016), Association for
  Computational Linguistics, Stroudsburg, PA, USA, 2016, pp. 144--153.
\newblock \href {https://doi.org/10.18653/v1/W16-2817}
  {\path{doi:10.18653/v1/W16-2817}}.

\bibitem{Addawood2016WhatMedia}
A.~A. Addawood, M.~N. Bashir, {What is Your Evidence? A Study of Controversial
  Topics on Social Media}, in: Proceedings of the 3rd Workshop on Argument
  Mining, Berlin, Germany, 2016, pp. 1--11.

\bibitem{Boltuzic2014BackDiscussions}
F.~Boltu{\v{z}}i{\'{c}}, J.~{\v{S}}najder, {Back up your Stance: Recognizing
  Arguments in Online Discussions}, in: Proceedings of the First Workshop on
  Argumentation Mining, Association for Computational Linguistics, Stroudsburg,
  PA, USA, 2014, pp. 49--58.
\newblock \href {https://doi.org/10.3115/v1/W14-2107}
  {\path{doi:10.3115/v1/W14-2107}}.

\bibitem{Kurtanovic2018OnEngineering}
Z.~Kurtanovi{\'{c}}, W.~Maalej, {On user rationale in software engineering},
  Requirements Engineering (2018) 1--23\href
  {https://doi.org/10.1007/s00766-018-0293-2}
  {\path{doi:10.1007/s00766-018-0293-2}}.

\bibitem{Park2014IdentifyingComments}
J.~Park, C.~Cardie, {Identifying Appropriate Support for Propositions in Online
  User Comments}, in: Proceedings of the First Workshop on Argumentation
  Mining, Baltimore, Maryland USA, 2014, pp. 29--38.

\bibitem{Park2015ConditionalComments}
J.~Park, A.~Katiyar, B.~Yang, {Conditional Random Fields for Identifying
  Appropriate Types of Support for Propositions in Online User Comments}, in:
  Proceedings of the 2nd Workshop on Argumentation Mining, Denver, Colorado,
  2015, pp. 39--44.

\bibitem{Rajendran2016ContextualReviews}
P.~Rajendran, {Contextual stance classification of opinions: A step towards
  enthymeme reconstruction in online reviews}, in: Proceedings of the 3rd
  Workshop on Argument Mining, Berlin, Germany, 2016, pp. 31--39.

\bibitem{Rajendran2018IsDataset}
P.~Rajendran, D.~Bollegala, S.~Parsons, {Is Something Better than Nothing?
  Automatically Predicting Stance-based Arguments using Deep Learning and Small
  Labelled Dataset}, in: 16th Annual Conference of the North American Chapter
  of the Association for Computational Linguistics: Human Language
  Technologies, New Orleans, Louisiana, 2018, pp. 28--34.

\bibitem{Belbachir2018UsingDetection}
F.~Belbachir, M.~Boughanem, {Using language models to improve opinion
  detection}, Information Processing {\&} Management 54~(6) (2018) 958--968.
\newblock \href {https://doi.org/10.1016/J.IPM.2018.07.001}
  {\path{doi:10.1016/J.IPM.2018.07.001}}.

\bibitem{Tubishat2018ImplicitChallenges}
M.~Tubishat, N.~Idris, M.~A. Abushariah, {Implicit aspect extraction in
  sentiment analysis: Review, taxonomy, oppportunities, and open challenges},
  Information Processing {\&} Management 54~(4) (2018) 545--563.
\newblock \href {https://doi.org/10.1016/J.IPM.2018.03.008}
  {\path{doi:10.1016/J.IPM.2018.03.008}}.

\bibitem{Mochales2011ArgumentationMining}
R.~Mochales, M.-F. Moens, {Argumentation mining}, Artificial Intelligence and
  Law 19~(1) (2011) 1--22.
\newblock \href {https://doi.org/10.1007/s10506-010-9104-x}
  {\path{doi:10.1007/s10506-010-9104-x}}.

\bibitem{Savelka2016ExtractingTerms}
J.~Savelka, K.~D. Ashley, {Extracting Case Law Sentences for Argumentation
  about the Meaning of Statutory Terms}, in: Proceedings of the 3rd Workshop on
  Argument Mining, 2016, pp. 50--59.

\bibitem{Lippi2018CLAUDETTE:Service}
M.~Lippi, P.~Palka, G.~Contissa, F.~Lagioia, H.-W. Micklitz, G.~Sartor,
  P.~Torroni, {CLAUDETTE: an Automated Detector of Potentially Unfair Clauses
  in Online Terms of Service}, arXiv preprint.

\bibitem{Green2018TowardsSchemes}
N.~L. Green,
  \href{http://www.medra.org/servlet/aliasResolver?alias=iospress&doi=10.3233/AAC-180038}{{Towards
  mining scientific discourse using argumentation schemes}}, Argument {\&}
  Computation 9~(2) (2018) 121--135.
\newblock \href {https://doi.org/10.3233/AAC-180038}
  {\path{doi:10.3233/AAC-180038}}.

\bibitem{Lauscher2018ArguminSci:Writing}
A.~Lauscher, G.~Glava{\v{s}}, K.~Eckert,
  \href{https://aclweb.org/anthology/papers/W/W18/W18-5203/}{{ArguminSci: A
  Tool for Analyzing Argumentation and Rhetorical Aspects in Scientific
  Writing}}, in: Proceedings of the 5th Workshop on Argument Mining,
  Association for Computational Linguistics, Brussels, Belgium, 2018, pp.
  22--28.

\bibitem{Lauscher2018AnPublications}
A.~Lauscher, G.~Glava{\v{s}}, S.~P. Ponzetto,
  \href{https://aclweb.org/anthology/papers/W/W18/W18-5206/}{{An
  Argument-Annotated Corpus of Scientific Publications}}, in: Proceedings of
  the 5th Workshop on Argument Mining, Association for Computational
  Linguistics, Brussels, Belgium, 2018, pp. 40--46.

\bibitem{Naderi2015ArgumentationDiscourse}
N.~Naderi, G.~Hirst, {Argumentation Mining in Parliamentary Discourse}, in:
  Workshop on Computational Models of Natural Argument, International Workshop
  on Empathic Computing, Springer, Cham, Bertinoro, Italy, 2015, pp. 16--25.
\newblock \href {https://doi.org/10.1007/978-3-319-46218-9{\_}2}
  {\path{doi:10.1007/978-3-319-46218-9{\_}2}}.

\bibitem{Bal2010TowardsEditorials}
B.~K. Bal, P.~S. Dizier, {Towards Building Annotated Resources for Analyzing
  Opinions and Argumentation in News Editorials}, in: Proceedings of the
  Seventh conference on International Language Resources and Evaluation,
  Valletta, Malta, 2010, pp. 1152--1158.

\bibitem{Sardianos2015ArgumentNews}
C.~Sardianos, I.~M. Katakis, G.~Petasis, V.~Karkaletsis, {Argument Extraction
  from News}, in: Proceedings of the 2nd Workshop on Argumentation Mining,
  Denver, Colorado, 2015, pp. 56--66.

\bibitem{REED2007ArgumentIntelligence}
C.~Reed, D.~Walton, F.~Macagno, {Argument diagramming in logic, law and
  artificial intelligence}, The Knowledge Engineering Review 22~(01) (2007) 87.
\newblock \href {https://doi.org/10.1017/S0269888907001051}
  {\path{doi:10.1017/S0269888907001051}}.

\bibitem{Skeppstedt2016UnsharedDebates}
M.~Skeppstedt, M.~Sahlgren, C.~Paradis, A.~Kerren, {Unshared task:
  (Dis)agreement in online debates}, in: Proceedings of the 3rd Workshop on
  Argument Mining, Berlin, Germany, 2016, pp. 154--159.

\bibitem{Freeman2011ArgumentTheory}
J.~B. Freeman, {Argument structure : representation and theory}, Springer,
  2011.

\bibitem{Walton2011HowIntelligence}
D.~Walton, {How to Refute an Argument Using Artificial Intelligence}, Studies
  in Logic, Grammar and Rhetoric 23~(36) (2011) 123--154.

\bibitem{Peldszus2016RhetoricalText}
A.~Peldszus, M.~Stede, {Rhetorical structure and argumentation structure in
  monologue text}, in: Proceedings of the 3rd Workshop on Argument Mining,
  Berlin, Germany, 2016, pp. 103--112.

\bibitem{Stab2014AnnotatingEssays}
C.~Stab, I.~Gurevych, {Annotating Argument Components and Relations in
  Persuasive Essays}, in: Proceedings of COLING 2014, the 25th International
  Conference on Computational Linguistics: Technical Papers ,, Dublin, Ireland,
  2014, pp. 1501--1510.

\bibitem{Green2017ManualMining}
N.~L. Green, {Manual Identification of Arguments with Implicit Conclusions
  Using Semantic Rules for Argument Mining}, in: Proceedings of the 4th
  Workshop on Argument Mining, Copenhagen, Denmark, 2017, pp. 73--78.

\bibitem{Boltuzic2016FillDebates}
F.~Boltu{\v{z}}ic, J.~Snajder, {Fill the Gap! Analyzing Implicit Premises
  between Claims from Online Debates}, in: Proceedings of the 3rd Workshop on
  Argument Mining, Berlin, Germany, 2016, pp. 124--133.

\bibitem{Habernal2018TheWarrants}
I.~Habernal, H.~Wachsmuth, I.~Gurevych, B.~Stein, {The Argument Reasoning
  Comprehension Task: Identification and Reconstruction of Implicit Warrants},
  in: 16th North American Chapter of the Association for Computational
  Linguistics: Human Language Technologies, New Orleans, Louisiana, USA, 2018,
  pp. 1930--1940.

\bibitem{MANN1988RhetoricalOrganization}
W.~C. Mann, S.~A. Thompson, {Rhetorical Structure Theory: Toward a functional
  theory of text organization}, Text - Interdisciplinary Journal for the Study
  of Discourse 8~(3) (1988) 243--281.
\newblock \href {https://doi.org/10.1515/text.1.1988.8.3.243}
  {\path{doi:10.1515/text.1.1988.8.3.243}}.

\bibitem{Reisert2015AArgumentation}
P.~Reisert, N.~Inoue, N.~Okazaki, K.~Inui, {A Computational Approach for
  Generating Toulmin Model Argumentation}, in: Proceedings of the 2nd Workshop
  on Argumentation Mining, Denver, Colorado, 2015, pp. 45--55.

\bibitem{Lee2018UnderstandingFacebook}
S.~Lee, T.~Ha, D.~Lee, J.~H. Kim, {Understanding the majority opinion formation
  process in online environments: An exploratory approach to Facebook},
  Information Processing {\&} Management 54~(6) (2018) 1115--1128.
\newblock \href {https://doi.org/10.1016/J.IPM.2018.08.002}
  {\path{doi:10.1016/J.IPM.2018.08.002}}.

\bibitem{Nguyen2018MultilingualEmbedding}
H.~T. Nguyen, M.~Le~Nguyen, {Multilingual opinion mining on YouTube – A
  convolutional N-gram BiLSTM word embedding}, Information Processing {\&}
  Management 54~(3) (2018) 451--462.
\newblock \href {https://doi.org/10.1016/J.IPM.2018.02.001}
  {\path{doi:10.1016/J.IPM.2018.02.001}}.

\bibitem{ChandraPandey2017TwitterMethod}
A.~Chandra~Pandey, D.~Singh~Rajpoot, M.~Saraswat, {Twitter sentiment analysis
  using hybrid cuckoo search method}, Information Processing {\&} Management
  53~(4) (2017) 764--779.
\newblock \href {https://doi.org/10.1016/J.IPM.2017.02.004}
  {\path{doi:10.1016/J.IPM.2017.02.004}}.

\bibitem{Giachanou2016LikeNot}
A.~Giachanou, F.~Crestani, {Like It or Not}, ACM Computing Surveys 49~(2)
  (2016) 1--41.
\newblock \href {https://doi.org/10.1145/2938640} {\path{doi:10.1145/2938640}}.

\bibitem{Peldszus2013FromTexts}
A.~Peldszus, M.~Stede, {From Argument Diagrams to Argumentation Mining in
  Texts}, International Journal of Cognitive Informatics and Natural
  Intelligence 7~(1) (2013) 1--31.
\newblock \href {https://doi.org/10.4018/jcini.2013010101}
  {\path{doi:10.4018/jcini.2013010101}}.

\bibitem{Lippi2016ArgumentationMining}
M.~Lippi, P.~Torroni, {Argumentation Mining}, ACM Transactions on Internet
  Technology 16~(2) (2016) 1--25.
\newblock \href {https://doi.org/10.1145/2850417} {\path{doi:10.1145/2850417}}.

\bibitem{Cabrio2018FiveAnalysis}
E.~Cabrio, S.~Villata, {Five Years of Argument Mining: a Data-driven Analysis},
  in: Proceedings of the Twenty-Seventh International Joint Conference on
  Artificial Intelligence, International Joint Conferences on Artificial
  Intelligence Organization, California, 2018, pp. 5427--5433.
\newblock \href {https://doi.org/10.24963/ijcai.2018/766}
  {\path{doi:10.24963/ijcai.2018/766}}.

\bibitem{Habernal2017ArgumentationDiscourse}
I.~Habernal, I.~Gurevych, {Argumentation Mining in User-Generated Web
  Discourse}, Computational Linguistics 43~(1) (2017) 125--179.
\newblock \href {https://doi.org/10.1162/COLI{\_}a{\_}00276}
  {\path{doi:10.1162/COLI{\_}a{\_}00276}}.

\bibitem{Palau2009ArgumentationMining}
R.~M. Palau, M.-F. Moens, {Argumentation mining}, in: Proceedings of the 12th
  International Conference on Artificial Intelligence and Law - ICAIL '09, ACM
  Press, New York, New York, USA, 2009, p.~98.
\newblock \href {https://doi.org/10.1145/1568234.1568246}
  {\path{doi:10.1145/1568234.1568246}}.

\bibitem{Aristotle2006OnDiscourse}
{Aristotle}, G.~A. Kennedy, {On Rhetoric: A Theory of Civic Discourse} (2006).

\bibitem{Walton2016ASchemes}
D.~Walton, F.~Macagno,
  \href{http://content.iospress.com/doi/1080/19462166.2015.1123772}{{A
  classification system for argumentation schemes}}, Argument {\&} Computation
  6~(3) (2016) 219--245.
\newblock \href {https://doi.org/10.1080/19462166.2015.1123772}
  {\path{doi:10.1080/19462166.2015.1123772}}.

\bibitem{Whately1857ElementsLogic.}
R.~Whately, {Elements of logic.}, Harper {\&} Brothers, New York, USA, 1857.

\bibitem{Beardsley1950PracticalLogic}
M.~C. Beardsley, {Practical Logic}, The Philosophical Quarterly\href
  {https://doi.org/10.2307/2216487} {\path{doi:10.2307/2216487}}.

\bibitem{Toulmin1958TheArgument}
S.~E. Toulmin, {The Uses of Argument}, Cambridge University Press, 1958.
\newblock \href {https://doi.org/10.1080/00048405985200191}
  {\path{doi:10.1080/00048405985200191}}.

\bibitem{Kuribayashi2018TowardsIdentification}
T.~Kuribayashi, P.~Reisert, N.~Inoue, K.~Inui, {Towards Exploiting
  Argumentative Context for Argumentative Relation Identification}, in:
  Proceedings of the 24th Annual Conference of the Society of Language
  Processing (March 2018), 2018, pp. 284--287.

\bibitem{Peldszus2015JointMining}
A.~Peldszus, M.~Stede, {Joint prediction in MST-style discourse parsing for
  argumentation mining}, in: Proceedings of the 2015 Conference on Empirical
  Methods in Natural Language Processing, Association for Computational
  Linguistics, Stroudsburg, PA, USA, 2015, pp. 938--948.
\newblock \href {https://doi.org/10.18653/v1/D15-1110}
  {\path{doi:10.18653/v1/D15-1110}}.

\bibitem{Freeman1991DialecticsStructure}
J.~B. Freeman, {Dialectics and the macrostructure of arguments: a theory of
  argument structure}, Foris Publications, 1991.

\bibitem{Wachsmuth2015SentimentArgumentation}
H.~Wachsmuth, J.~Kiesel, B.~Stein, {Sentiment Flow - A General Model of Web
  Review Argumentation}, in: Proceedings of the 2015 Conference on Empirical
  Methods in Natural Language Processing, Lisbon, Portugal, 2015, pp. 601--611.

\bibitem{Mann1984DiscourseGeneration}
W.~C. Mann, {Discourse structures for text generation}, in: Proceedings of the
  10th international conference on Computational linguistics -, Association for
  Computational Linguistics, Morristown, NJ, USA, 1984, pp. 367--375.
\newblock \href {https://doi.org/10.3115/980431.980567}
  {\path{doi:10.3115/980431.980567}}.

\bibitem{Carstens2017UsingProblems}
L.~Carstens, F.~Toni, {Using Argumentation to Improve Classification in Natural
  Language Problems}, ACM Transactions on Internet Technology 17~(3) (2017)
  1--23.
\newblock \href {https://doi.org/10.1145/3017679} {\path{doi:10.1145/3017679}}.

\bibitem{Pollock1987DefeasibleReasoning}
J.~L. Pollock, {Defeasible reasoning}, Cognitive Science 11~(4) (1987)
  481--518.
\newblock \href {https://doi.org/10.1016/S0364-0213(87)80017-4}
  {\path{doi:10.1016/S0364-0213(87)80017-4}}.

\bibitem{Dung1995OnGames}
P.~M. Dung, {On the acceptability of arguments and its fundamental role in
  nonmonotonic reasoning, logic programming and n-person games}, Artificial
  Intelligence 77~(2) (1995) 321--357.
\newblock \href {https://doi.org/10.1016/0004-3702(94)00041-X}
  {\path{doi:10.1016/0004-3702(94)00041-X}}.

\bibitem{Krause1995AUncertainty}
P.~Krause, S.~Ambler, M.~Elvang-Goransson, J.~Fox, {A Logic of Argumentation
  for Reasoning under Uncertainty}, Computational Intelligence 11~(1) (1995)
  113--131.
\newblock \href {https://doi.org/10.1111/j.1467-8640.1995.tb00025.x}
  {\path{doi:10.1111/j.1467-8640.1995.tb00025.x}}.

\bibitem{Pollock2001DefeasibleJustification}
J.~L. Pollock, {Defeasible reasoning with variable degrees of justification},
  Artificial Intelligence 133~(1-2) (2001) 233--282.
\newblock \href {https://doi.org/10.1016/S0004-3702(01)00145-X}
  {\path{doi:10.1016/S0004-3702(01)00145-X}}.

\bibitem{Parsons1996NeogotiationReport}
S.~D. Parsons, N.~R. Jennings, {Neogotiation Through Argumentation - A
  Preliminary Report}, in: 2nd Int. Conf. on Multi-Agent Systems, Japan, 1996,
  pp. 267--274.

\bibitem{Jennings2001AutomatedChallenges}
N.~Jennings, P.~Faratin, A.~Lomuscio, S.~Parsons, M.~Wooldridge, C.~Sierra,
  {Automated Negotiation: Prospects, Methods and Challenges}, Group Decision
  and Negotiation 10~(2) (2001) 199--215.
\newblock \href {https://doi.org/10.1023/A:1008746126376}
  {\path{doi:10.1023/A:1008746126376}}.

\bibitem{GRASSO2000DialecticalNutrition}
F.~Grasso, A.~Cawsey, R.~Jones, {Dialectical argumentation to solve conflicts
  in advice giving: a case study in the promotion of healthy nutrition},
  International Journal of Human-Computer Studies 53~(6) (2000) 1077--1115.
\newblock \href {https://doi.org/10.1006/IJHC.2000.0429}
  {\path{doi:10.1006/IJHC.2000.0429}}.

\bibitem{Stab2018Cross-topicNetworks}
C.~Stab, T.~Miller, I.~Gurevych,
  \href{https://dblp.uni-trier.de/rec/bibtex/journals/corr/abs-1802-05758}{{Cross-topic
  Argument Mining from Heterogeneous Sources Using Attention-based Neural
  Networks}}, CoRR.

\bibitem{Addawood2017StanceCase}
A.~Addawood, J.~Schneider, M.~Bashir, {Stance Classification of Twitter
  Debates: The Encryption Debate as A Use Case}, in: Proceedings of the 8th
  International Conference on Social Media {\&} Society, ACM Press, New York,
  New York, USA, 2017, pp. 1--10.
\newblock \href {https://doi.org/10.1145/3097286.3097288}
  {\path{doi:10.1145/3097286.3097288}}.

\bibitem{Bosc2016TweetiesMedia.}
T.~Bosc, E.~Cabrio, S.~Villata, {Tweeties Squabbling: Positive and Negative
  Results in Applying Argument Mining on Social Media.}, in: Proceedings of the
  6th International Conference on Computational Models of Argument, Potsdam,
  Germany, 2016, pp. 21--32.

\bibitem{Deturck2018ERTIMMC2:Retrieval}
K.~Deturck, D.~Nouvel, F.~Segond, {ERTIM@MC2: Diversified Argumentative Tweets
  Retrieval}, in: CLEF MC2 2018 Lab Overview, Avignon, France, 2018, pp.
  302--308.

\bibitem{Dusmanu2017ArgumentSources}
M.~Dusmanu, E.~Cabrio, S.~Villata, {Argument Mining on Twitter: Arguments,
  Facts and Sources}, in: Proceedings of the 2017 Conference on Empirical
  Methods in Natural Language Processing, Copenhagen, Denmark, 2017, pp.
  2317--2322.

\bibitem{Sendi2018OpinionModeling.}
S.~Sendi, C.~Latiri, {Opinion Argumentation based on Combined Information
  Retrieval and Topic Modeling.}, in: Working Notes of CLEF 2018 - Conference
  and Labs of the Forum, Avignon, France, 2018.

\bibitem{Dufour2018LIACLEFNetwork.}
R.~Dufour, R.~Mickael, A.~Delorme, D.~Malinas, {LIA@CLEF 2018: Mining Events
  Opinion Argumentation from Raw Unlabeled Twitter Data using Convolutional
  Neural Network.}, in: Working Notes of CLEF 2018 - Conference and Labs of the
  Evaluation Forum, Avignon, France, 2018.

\bibitem{Cocarascu2018CombiningDatasets}
O.~Cocarascu, F.~Toni, {Combining deep learning and argumentative reasoning for
  the analysis of social media textual content using small datasets},
  Computational Linguistics (2018) 1--37\href
  {https://doi.org/10.1162/coli{\_}a{\_}00338}
  {\path{doi:10.1162/coli{\_}a{\_}00338}}.

\bibitem{Ma2018ClaimTwitter}
W.~Ma, W.~Chao, Z.~Luo, X.~Jiang, {Claim Retrieval in Twitter}, in: Web
  Information Systems Engineering – WISE 2018, Dubai, United Arab Emirates,
  2018, pp. 297--307.
\newblock \href {https://doi.org/10.1007/978-3-030-02922-7{\_}20}
  {\path{doi:10.1007/978-3-030-02922-7{\_}20}}.

\bibitem{Mohammad2017StanceTweets}
S.~M. Mohammad, P.~Sobhani, S.~Kiritchenko, {Stance and Sentiment in Tweets},
  ACM Transactions on Internet Technology 17~(3) (2017) 1--23.
\newblock \href {https://doi.org/10.1145/3003433} {\path{doi:10.1145/3003433}}.

\bibitem{Zarrella2016MITREDetection}
G.~Zarrella, A.~Marsh, {MITRE at SemEval-2016 Task 6: Transfer Learning for
  Stance Detection}, in: International Workshop on Semantic Evaluation
  (SemEval-2016), San Diego, California, 2016, p. 458–463.

\bibitem{Wei2018Multi-TargetNetwork}
P.~Wei, J.~Lin, W.~Mao, {Multi-Target Stance Detection via a Dynamic
  Memory-Augmented Network}, in: The 41st International ACM SIGIR Conference on
  Research {\&} Development in Information Retrieval - SIGIR '18, ACM Press,
  New York, New York, USA, 2018, pp. 1229--1232.
\newblock \href {https://doi.org/10.1145/3209978.3210145}
  {\path{doi:10.1145/3209978.3210145}}.

\bibitem{Ebrahimi2016WeaklyBootstrapping}
J.~Ebrahimi, D.~Dou, D.~Lowd, {Weakly Supervised Tweet Stance Classification by
  Relational Bootstrapping}, in: Proceedings of the 2016 Conference on
  Empirical Methods in Natural Language Processing, Austin, Texas, 2016, p.
  1012–1017.

\bibitem{Lai2018StanceDebate}
M.~Lai, V.~Patti, G.~Ruffo, P.~Rosso, {Stance Evolution and Twitter
  Interactions in an Italian Political Debate}, in: NLDB 2018: Natural Language
  Processing and Information Systems, Springer, Cham, Paris, France, 2018, pp.
  15--27.
\newblock \href {https://doi.org/10.1007/978-3-319-91947-8{\_}2}
  {\path{doi:10.1007/978-3-319-91947-8{\_}2}}.

\bibitem{Johnson2016IdentifyingTwitter}
K.~Johnson, D.~Goldwasser, {Identifying Stance by Analyzing Political Discourse
  on Twitter}, in: Proceedings of the First Workshop on NLP and Computational
  Social Science, Association for Computational Linguistics, Stroudsburg, PA,
  USA, 2016, pp. 66--75.
\newblock \href {https://doi.org/10.18653/v1/W16-5609}
  {\path{doi:10.18653/v1/W16-5609}}.

\bibitem{Konstantinovskiy2018TowardsDetection}
L.~Konstantinovskiy, O.~Price, M.~Babakar, A.~Zubiaga, {Towards Automated
  Factchecking: Developing an Annotation Schema and Benchmark for Consistent
  Automated Claim Detection}, in: EMNLP 2018: Conference on Empirical Methods
  in Natural Language Processing, Brussels, Belgium, 2018.

\bibitem{Carletta1996AssessingStatistic}
J.~Carletta, \href{https://dl.acm.org/citation.cfm?id=230390}{{Assessing
  agreement on classification tasks: the kappa statistic}}, Computational
  Linguistics 22~(2) (1996) 249--254.

\bibitem{krippendorff2004MeasuringData}
K.~krippendorff,
  \href{http://link.springer.com/10.1007/s11135-004-8107-7}{{Measuring the
  Reliability of Qualitative Text Analysis Data}}, Quality {\&} Quantity 38~(6)
  (2004) 787--800.
\newblock \href {https://doi.org/10.1007/s11135-004-8107-7}
  {\path{doi:10.1007/s11135-004-8107-7}}.

\bibitem{Dice1945MeasuresSpecies}
L.~R. Dice, \href{http://doi.wiley.com/10.2307/1932409}{{Measures of the Amount
  of Ecologic Association Between Species}}, Ecology 26~(3) (1945) 297--302.
\newblock \href {https://doi.org/10.2307/1932409} {\path{doi:10.2307/1932409}}.

\bibitem{Fleiss1971MeasuringRaters.}
J.~L. Fleiss, \href{http://content.apa.org/journals/bul/76/5/378}{{Measuring
  nominal scale agreement among many raters.}}, Psychological Bulletin 76~(5)
  (1971) 378--382.
\newblock \href {https://doi.org/10.1037/h0031619}
  {\path{doi:10.1037/h0031619}}.

\bibitem{Bosc2016DART:Scholar}
T.~Bosc, E.~Cabrio, S.~Villata, {DART: a Dataset of Arguments and their
  Relations on Twitter - Semantic Scholar}, in: LREC, Portoro{\v{z}}, Slovenia,
  2016, pp. 1258--1263.

\bibitem{Pennebaker1996CognitiveDisclosure}
J.~W. Pennebaker, M.~E. Francis, {Cognitive, Emotional, and Language Processes
  in Disclosure}, Cognition and Emotion 10~(6) (1996) 601--626.
\newblock \href {https://doi.org/10.1080/026999396380079}
  {\path{doi:10.1080/026999396380079}}.

\bibitem{Pennebaker1997WritingProcess}
J.~W. Pennebaker, {Writing About Emotional Experiences as a Therapeutic
  Process}, Psychological Science 8~(3) (1997) 162--166.
\newblock \href {https://doi.org/10.1111/j.1467-9280.1997.tb00403.x}
  {\path{doi:10.1111/j.1467-9280.1997.tb00403.x}}.

\bibitem{Guo2013LinkingMedia}
W.~Guo, H.~Li, H.~Ji, M.~Diab, {Linking Tweets to News: A Framework to Enrich
  Short Text Data in Social Media}, in: Proceedings of the 51st Annual Meeting
  of the Association for Computational Linguists, Sofia, Bulgaria, 2013, p.
  239–249.

\bibitem{Tan2017SpotTwitter.}
S.~Tan, {Spot the lie: Detecting untruthful online opinion on twitter.}, Ph.D.
  thesis, Imperial College London (2017).

\bibitem{Goudas2014ArgumentMedia}
T.~Goudas, C.~Louizos, G.~Petasis, V.~Karkaletsis, {Argument Extraction from
  News, Blogs, and Social Media}, in: Artificial Intelligence: Methods and
  Applications.SETN 2014., Springer, Cham, 2014, pp. 287--299.
\newblock \href {https://doi.org/10.1007/978-3-319-07064-3{\_}23}
  {\path{doi:10.1007/978-3-319-07064-3{\_}23}}.

\bibitem{Roitman2016OnTopics}
H.~Roitman, S.~Hummel, E.~Rabinovich, B.~Sznajder, N.~Slonim, E.~Aharoni, {On
  the Retrieval of Wikipedia Articles Containing Claims on Controversial
  Topics}, in: Proceedings of the 25th International Conference Companion on
  World Wide Web - WWW '16 Companion, ACM Press, New York, New York, USA, 2016,
  pp. 991--996.
\newblock \href {https://doi.org/10.1145/2872518.2891115}
  {\path{doi:10.1145/2872518.2891115}}.

\bibitem{Lawrence2017DebatingArgument}
J.~Lawrence, M.~Snaith, B.~Konat, K.~Budzynska, C.~Reed, {Debating Technology
  for Dialogical Argument}, ACM Transactions on Internet Technology 17~(3)
  (2017) 1--23.
\newblock \href {https://doi.org/10.1145/3007210} {\path{doi:10.1145/3007210}}.

\bibitem{Morio2018AnnotatingMining}
G.~Morio, K.~Fujita, {Annotating Online Civic Discussion Threads for Argument
  Mining}, in: Proceedings of the IEEE/WIC/ACM International Conference on Web
  Intelligence 2018 (WI’18), IEEE, Santiago, Chile, 2018, pp. 801--807.
\newblock \href {https://doi.org/10.1109/IIAI-AAI.2017.123}
  {\path{doi:10.1109/IIAI-AAI.2017.123}}.

\bibitem{Galassi2018ArgumentativeLearning}
A.~Galassi, M.~Lippi, P.~Torroni, {Argumentative Link Prediction using Residual
  Networks and Multi-Objective Learning}, in: Proceedings of the 5th Workshop
  on Argument Mining, Brussels, Belgium, 2018, pp. 1--10.

\bibitem{Eidelman2019ArgumentERulemaking}
V.~Eidelman, B.~Grom, \href{http://arxiv.org/abs/1905.00572
  http://arxiv.org/abs/1905.00572}{{Argument Identification in Public Comments
  from eRulemaking}} (5 2019).
\newblock \href {https://doi.org/10.1145/3322640.3326714}
  {\path{doi:10.1145/3322640.3326714}}.

\bibitem{Lai2017ExtractingOpinion}
M.~Lai, M.~Tambuscio, V.~Patti, G.~Ruffo, P.~Rosso,
  \href{http://link.springer.com/10.1007/978-3-319-65813-1_10}{{Extracting
  Graph Topological Information and Users’ Opinion}}, Springer, Cham, 2017,
  pp. 112--118.
\newblock \href {https://doi.org/10.1007/978-3-319-65813-1{\_}10}
  {\path{doi:10.1007/978-3-319-65813-1{\_}10}}.

\bibitem{Maynard2017AAnalysis}
D.~Maynard, I.~Roberts, M.~A. Greenwood, D.~Rout, K.~Bontcheva, {A framework
  for real-time semantic social media analysis}, Web Semantics: Science,
  Services and Agents on the World Wide Web 44 (2017) 75--88.
\newblock \href {https://doi.org/10.1016/J.WEBSEM.2017.05.002}
  {\path{doi:10.1016/J.WEBSEM.2017.05.002}}.

\bibitem{Cortis2017SemEval-2017News}
K.~Cortis, A.~Freitas, T.~Daudert, M.~H{\"{u}}rlimann, M.~Zarrouk,
  S.~Handschuh, B.~Davis,
  \href{http://www.aclweb.org/anthology/S17-2089}{{SemEval-2017 Task 5:
  Fine-Grained Sentiment Analysis on Financial Microblogs and News}}, in:
  Proceedings of the 11th International Workshop on Semantic Evaluations
  (SemEval-2017), Vancouver, Canad, 2017, pp. 519--535.

\bibitem{Schulz2018Multi-TaskSettings}
C.~Schulz, S.~Eger, J.~Daxenberger, T.~Kahse, I.~Gurevych, {Multi-Task Learning
  for Argumentation Mining in Low-Resource Settings}, in: NAACL HLT 2018, 2018,
  pp. 35--41.

\bibitem{Lawrence2017UsingERulemaking}
J.~Lawrence, J.~Park, K.~Budzynska, C.~Cardie, B.~Konat, C.~Reed, {Using
  Argumentative Structure to Interpret Debates in Online Deliberative Democracy
  and eRulemaking}, ACM Transactions on Internet Technology 17~(3) (2017)
  1--22.
\newblock \href {https://doi.org/10.1145/3032989} {\path{doi:10.1145/3032989}}.

\bibitem{Grcar2017StanceReferendum}
M.~Gr{\v{c}}ar, D.~Cherepnalkoski, I.~Mozeti{\v{c}}, P.~Kralj~Novak, {Stance
  and influence of Twitter users regarding the Brexit referendum},
  Computational Social Networks 4~(1) (2017) 6.
\newblock \href {https://doi.org/10.1186/s40649-017-0042-6}
  {\path{doi:10.1186/s40649-017-0042-6}}.

\bibitem{Lytos2018ArgumentationKnowledge}
A.~Lytos, T.~Lagkas, P.~Sarigiannidis, K.~Bontcheva,
  \href{https://www.researchgate.net/publication/327728501}{{Argumentation
  Mining: Exploiting Multiple Sources and Background Knowledge}}, in: 12th
  South East European Doctoral Student Conference DSC2018, 2018.

\bibitem{Liu2012SentimentMining}
B.~Liu, {Sentiment Analysis and Opinion Mining}, Morgan {\&} Claypool
  Publishers, 2012.

\bibitem{Wei2016PkudblabDetection}
W.~Wei, X.~Zhang, X.~Liu, W.~Chen, T.~Wang, {pkudblab at SemEval-2016 Task 6 :
  A Specific Convolutional Neural Network System for Effective Stance
  Detection}, in: Proceedings of SemEval-2016, San Diego, California, 2016, pp.
  384--388.

\bibitem{Mohammad2016SemEval-2016Tweets}
S.~M. Mohammad, S.~Kiritchenko, P.~Sobhani, X.~Zhu, C.~Cherry, {SemEval-2016
  Task 6: Detecting Stance in Tweets}, in: International Workshop on Semantic
  Evaluation (SemEval-2016), San Diego, California, 2016, pp. 31--41.

\bibitem{Rinott2015ShowDetection}
R.~Rinott, L.~Dankin, C.~Alzate~Perez, M.~M. Khapra, E.~Aharoni, N.~Slonim,
  {Show Me Your Evidence - an Automatic Method for Context Dependent Evidence
  Detection}, in: Proceedings of the 2015 Conference on Empirical Methods in
  Natural Language Processing, Association for Computational Linguistics,
  Stroudsburg, PA, USA, 2015, pp. 440--450.
\newblock \href {https://doi.org/10.18653/v1/D15-1050}
  {\path{doi:10.18653/v1/D15-1050}}.

\bibitem{Levy2014ContextDetection}
R.~Levy, Y.~Bilu, D.~Hershcovich, E.~Aharoni, N.~Slonim, {Context Dependent
  Claim Detection}, in: COLING - International Committee on Computational
  Linguistics, Dublin, Ireland, 2014, pp. 1489--1500.

\bibitem{Stenetorp2012BRAT:Annotation}
P.~Stenetorp, S.~Pyysalo, G.~Topi, T.~Ohta, S.~Ananiadou, J.~i. Tsujii, {BRAT:
  a Web-based Tool for NLP-Assisted Text Annotation}, in: Proceedings of the
  13th Conference of the European Chapter of the Association for Computational
  Linguistics, Avignon, France, 2012, pp. 102--107.

\bibitem{Stenetorp2011BioNLPResources}
P.~Stenetorp, G.~Topi{\'{c}}, S.~Pyysalo, T.~Ohta, J.-D. Kim, J.~Tsujii,
  {BioNLP Shared Task 2011: supporting resources}, in: Proceedings of the
  BioNLP Shared Task 2011 Workshop, Association for Computational Linguistics,
  Portland, Oregon, 2011, pp. 112--120.

\bibitem{EckartDeCastilho2016AStructures}
R.~Eckart De~Castilho, E.~Ujdricza-Maydt, S.~M. Yimam, S.~Hartmann,
  I.~Gurevych, A.~Frank, C.~Biemann, {A Web-based Tool for the Integrated
  Annotation of Semantic and Syntactic Structures}, in: Proceedings of the
  Workshop on Language Technology Resources and Tools for Digital Humanities
  (LT4DH), Osaka, Japan, 2016, pp. 76--84.

\bibitem{Sonntag2014GraPAT:Annotations}
J.~Sonntag, M.~Stede, {GraPAT: a Tool for Graph Annotations}, in: Proceedings
  of the Ninth International Conference on Language Resources and Evaluation
  (LREC-2014), Reykjavik, Iceland, 2014, pp. 4141--4151.

\bibitem{ODonnell2000RSTToolTheory}
M.~O'Donnell, {RSTTool 2.4 -- A Markup Tool for Rhetorical Structure Theory},
  in: Proceedings of the International Natural Language Generation Conference
  (INLG'2000), Mitzpe Ramon, Israel, 2000, pp. 253 -- 256.

\bibitem{Cunningham2013GettingAnalytics}
H.~Cunningham, V.~Tablan, A.~Roberts, K.~Bontcheva, {Getting More Out of
  Biomedical Documents with GATE's Full Lifecycle Open Source Text Analytics},
  PLoS Computational Biology 9~(2) (2013) e1002854.
\newblock \href {https://doi.org/10.1371/journal.pcbi.1002854}
  {\path{doi:10.1371/journal.pcbi.1002854}}.

\bibitem{Bontcheva2013GATEFramework}
K.~Bontcheva, H.~Cunningham, I.~Roberts, A.~Roberts, V.~Tablan, N.~Aswani,
  G.~Gorrell, {GATE Teamware: a web-based, collaborative text annotation
  framework}, Language Resources and Evaluation 47~(4) (2013) 1007--1029.
\newblock \href {https://doi.org/10.1007/s10579-013-9215-6}
  {\path{doi:10.1007/s10579-013-9215-6}}.

\bibitem{Kirschner2015LinkingPublications}
C.~Kirschner, J.~Eckle-Kohler, I.~Gurevych, {Linking the Thoughts: Analysis of
  Argumentation Structures in Scientific Publications}, in: Proceedings of the
  2nd Workshop on Argumentation Mining, Denver, Colorado, 2015, pp. 1--11.

\bibitem{Saint-Dizier2012ProcessingPlatform}
P.~Saint-Dizier, {Processing natural language arguments with the TextCoop
  platform}, Argument {\&} Computation 3~(1) (2012) 49--82.
\newblock \href {https://doi.org/10.1080/19462166.2012.663539}
  {\path{doi:10.1080/19462166.2012.663539}}.

\bibitem{Eryiit2013TURKSENT:Media}
U.~Eryiit, F.~Samet, C.~Etin, M.~Yanık, T.~Temel, {TURKSENT: A Sentiment
  Annotation Tool for Social Media}, in: Proceedings of the 7th Linguistic
  Annotation Workshop {\&} Interoperability with Discourse, Sofia, Bulgaria,
  2013, pp. 131--134.

\bibitem{Yimam2013WebAnno:Annotations}
S.~M. Yimam, I.~Gurevych, R.~Eckart De~Castilho, C.~Biemann, {WebAnno: A
  Flexible, Web-based and Visually Supported System for Distributed
  Annotations}, in: Proceedings of the 51st Annual Meeting of the Association
  for Computational Linguistics, Sofia, Bulgaria, 2013, pp. 1--6.

\bibitem{Lippi2016MARGOT:Mining}
M.~Lippi, P.~Torroni, {MARGOT: A web server for argumentation mining}, Expert
  Systems with Applications 65 (2016) 292--303.
\newblock \href {https://doi.org/10.1016/J.ESWA.2016.08.050}
  {\path{doi:10.1016/J.ESWA.2016.08.050}}.

\bibitem{JanierMathilde2014OVA+:Interface}
{Janier Mathilde}, {Lawrence John}, {Reed Chris}, {OVA+: An argument analysis
  interface}, in: Proceedings of the 5th International Conference on
  Computational Models of Argument (COMMA’14)., 2014, p. 463–464.

\bibitem{SnaithMark2012TOAST:Implementation}
{Snaith Mark}, {Reed Chris}, {TOAST: Online ASPIC+ implementation}, in:
  Proceedings of the Fourth International Conference on Computational Models of
  Argument (COMMA 2012), Vienna, Austria, 2012.

\bibitem{Wachsmuth2017BuildingWeb}
H.~Wachsmuth, M.~Potthast, K.~Al~Khatib, Y.~Ajjour, J.~Puschmann, J.~Qu,
  J.~Dorsch, V.~Morari, J.~Bevendorff, B.~Stein, {Building an Argument Search
  Engine for the Web}, in: Proceedings of the 4th Workshop on Argument Mining,
  Association for Computational Linguistics, Stroudsburg, PA, USA, 2017, pp.
  49--59.
\newblock \href {https://doi.org/10.18653/v1/W17-5106}
  {\path{doi:10.18653/v1/W17-5106}}.

\bibitem{Stab2018ArgumenText:Sources}
C.~Stab, J.~Daxenberger, C.~Stahlhut, T.~Miller, B.~Schiller, C.~Tauchmann,
  S.~Eger, I.~Gurevych, {ArgumenText: Searching for Arguments in Heterogeneous
  Sources}, in: Proceedings of the 2018 Conference of the North American
  Chapter of the Association for Computational Linguistics: Demonstrations,
  Association for Computational Linguistics, Stroudsburg, PA, USA, 2018, pp.
  21--25.
\newblock \href {https://doi.org/10.18653/v1/N18-5005}
  {\path{doi:10.18653/v1/N18-5005}}.

\bibitem{vanGelder2007TheRationaleTM}
T.~van Gelder, {The rationale for RationaleTM}, Law, Probability and Risk
  6~(1-4) (2007) 23--42.
\newblock \href {https://doi.org/10.1093/lpr/mgm032}
  {\path{doi:10.1093/lpr/mgm032}}.

\bibitem{Manning2014TheToolkit}
C.~Manning, M.~Surdeanu, J.~Bauer, J.~Finkel, S.~Bethard, D.~McClosky, {The
  Stanford CoreNLP Natural Language Processing Toolkit}, in: Proceedings of
  52nd Annual Meeting of the Association for Computational Linguistics: System
  Demonstrations, Association for Computational Linguistics, Baltimore,
  Maryland, 2014, pp. 55--60.
\newblock \href {https://doi.org/10.3115/v1/P14-5010}
  {\path{doi:10.3115/v1/P14-5010}}.

\bibitem{REED2004Araucaria:Representation}
C.~Reed, G.~Rowe, {Araucaria: Software For Argument Analysis, Diagramming And
  Representation}, International Journal on Artificial Intelligence Tools
  13~(04) (2004) 961--979.
\newblock \href {https://doi.org/10.1142/S0218213004001922}
  {\path{doi:10.1142/S0218213004001922}}.

\bibitem{Snaith2010MixedDeliberation}
M.~Snaith, J.~Lawrence, C.~Reed, {Mixed Initiative Argument in Public
  Deliberation}, in: International Conference on Online Deliberation, Leeds,
  UK, 2010, pp. 2--13.

\bibitem{Reed2007DialogicalDebates}
C.~Reed, S.~Wells, {Dialogical Argument as an Interface to Complex Debates},
  IEEE Intelligent Systems 22~(6) (2007) 60--65.
\newblock \href {https://doi.org/10.1109/MIS.2007.106}
  {\path{doi:10.1109/MIS.2007.106}}.

\bibitem{Reed2017TheArgumentation}
C.~Reed, K.~Budzynska, R.~Duthie, M.~Janier, B.~Konat, J.~Lawrence, A.~Pease,
  M.~Snaith, {The Argument Web: an Online Ecosystem of Tools, Systems and
  Services for Argumentation}, Philosophy {\&} Technology 30~(2) (2017)
  137--160.
\newblock \href {https://doi.org/10.1007/s13347-017-0260-8}
  {\path{doi:10.1007/s13347-017-0260-8}}.

\bibitem{Bex2014ArguBlogging:Web}
F.~Bex, M.~Snaith, J.~Lawrence, C.~Reed, {ArguBlogging: An application for the
  Argument Web}, Web Semantics: Science, Services and Agents on the World Wide
  Web 25 (2014) 9--15.
\newblock \href {https://doi.org/10.1016/J.WEBSEM.2014.02.002}
  {\path{doi:10.1016/J.WEBSEM.2014.02.002}}.

\bibitem{Lippi2015Context-independentMining}
M.~Lippi, P.~Torroni, {Context-independent claim detection for argument
  mining}, in: Proceedings of the 24th International Conference on Artificial
  Intelligence, AAAI Press = The Association for the Advancement of Artificial
  Intelligence Press, Buenos Aires, Argentina, 2015, pp. 185--191.

\bibitem{Sbarski2008VisualizingStructure}
P.~Sbarski, T.~van Gelder, K.~Marriott, D.~Prager, A.~Bulka, {Visualizing
  Argument Structure}, in: International Symposium on Visual Computing 2008:
  Visualizing Argument Structure, Springer, Berlin, Heidelberg, Las Vegas, NV,
  USA, 2008, pp. 129--138.
\newblock \href {https://doi.org/10.1007/978-3-540-89639-5{\_}13}
  {\path{doi:10.1007/978-3-540-89639-5{\_}13}}.

\bibitem{Marriott2011Hi-TreesLayout}
K.~Marriott, P.~Sbarski, T.~van Gelder, D.~Prager, A.~Bulka, {Hi-Trees and
  Their Layout}, IEEE Transactions on Visualization and Computer Graphics
  17~(3) (2011) 290--304.
\newblock \href {https://doi.org/10.1109/TVCG.2010.45}
  {\path{doi:10.1109/TVCG.2010.45}}.

\bibitem{Gelder2000LearningApproach}
T.~V. Gelder, {Learning to reason: a Reason!-Able approach}, in: Cognitive
  Science in Australia, 2000: Proceedings of the Fifth Australasian Cognitive
  Science Society Conference, Adelaide, 2000.

\bibitem{Al-Khatib2016Cross-DomainSupervision}
K.~Al-Khatib, H.~Wachsmuth, M.~Hagen, J.~K{\"{o}}hler, B.~Stein, {Cross-Domain
  Mining of Argumentative Text through Distant Supervision}, in: Proceedings of
  NAACL-HLT 2016, San Diego, California, 2016, pp. 1395--1404.

\bibitem{Wachsmuth2017AAnalysis}
H.~Wachsmuth, B.~Stein, {A Universal Model for Discourse-Level Argumentation
  Analysis}, ACM Transactions on Internet Technology 17~(3) (2017) 1--24.
\newblock \href {https://doi.org/10.1145/2957757} {\path{doi:10.1145/2957757}}.

\bibitem{Karimi2018NewsAhead}
M.~Karimi, D.~Jannach, M.~Jugovac, {News recommender systems – Survey and
  roads ahead}, Information Processing {\&} Management 54~(6) (2018)
  1203--1227.
\newblock \href {https://doi.org/10.1016/J.IPM.2018.04.008}
  {\path{doi:10.1016/J.IPM.2018.04.008}}.

\bibitem{Boltuzic2015IdentifyingSimilarity}
F.~Boltu{\v{z}}i{\'{c}}, J.~{\v{S}}najder, {Identifying Prominent Arguments in
  Online Debates Using Semantic Textual Similarity}, in: 2nd Workshop on
  Argumentation Mining (ARG-MINING 2015), Denver, Colorado, USA, 2015, pp.
  110--115.

\bibitem{Duan2019AnDetection}
X.~Duan, M.~Liao, X.~Zhao, W.~Wu, P.~Lv,
  \href{http://link.springer.com/10.1007/978-981-13-7983-3_18}{{An Unsupervised
  Joint Model for Claim Detection}}, in: International Conference on Cognitive
  Systems and Signal Processing: Cognitive Systems and Signal Processing,
  Springer, Singapore, 2019, pp. 197--209.
\newblock \href {https://doi.org/10.1007/978-981-13-7983-3{\_}18}
  {\path{doi:10.1007/978-981-13-7983-3{\_}18}}.

\bibitem{Shnarch2018WillMining}
E.~Shnarch, C.~Alzate, L.~Dankin, M.~Gleize, Y.~Hou, L.~Choshen, R.~Aharonov,
  N.~Slonim, \href{https://aclanthology.info/papers/P18-2095/p18-2095}{{Will it
  Blend? Blending Weak and Strong Labeled Data in a Neural Network for
  Argumentation Mining}}, in: Proceedings of the 56th Annual Meeting of the
  Association for Computational Linguistics (Volume 2: Short Papers), Vol.~2,
  Melbourne, Australia, 2018, pp. 599--605.

\bibitem{Silva2016ALearning}
N.~F. F.~D. Silva, L.~F.~S. Coletta, E.~R. Hruschka, {A Survey and Comparative
  Study of Tweet Sentiment Analysis via Semi-Supervised Learning}, ACM
  Computing Surveys 49~(1) (2016) 1--26.
\newblock \href {https://doi.org/10.1145/2932708} {\path{doi:10.1145/2932708}}.

\bibitem{Day2017DeepReview}
M.-Y. Day, Y.-D. Lin, {Deep Learning for Sentiment Analysis on Google Play
  Consumer Review}, in: 2017 IEEE International Conference on Information Reuse
  and Integration (IRI), IEEE, San Diego, CA, USA, 2017, pp. 382--388.
\newblock \href {https://doi.org/10.1109/IRI.2017.79}
  {\path{doi:10.1109/IRI.2017.79}}.

\bibitem{Wallace2015SparseSentiment}
B.~C. Wallace, D.~K. Choe, E.~Charniak, {Sparse, contextually informed models
  for irony detection: Exploiting user communities, entities and sentiment},
  in: 53rd Annual Meeting of the Association for Computational Linguistics and
  the 7th International Joint Conference on Natural Language Processing of the
  Asian Federation of Natural Language Processing, Proceedings of the
  Conference, Association for Computational Linguistics (ACL), Beijing, China,
  2015, pp. 1035--1044.

\bibitem{Wallace2014HumansToo}
B.~C. Wallace, D.~K. Choe, L.~Kertz, E.~Charniak, {Humans require context to
  infer ironic intent (so computers probably do, too)}, in: 52nd Annual Meeting
  of the Association for Computational Linguistics, Association for
  Computational Linguistics (ACL), 2014, pp. 512--516.

\end{thebibliography}

\end{document}